\documentclass[a4paper,11pt]{article}

\usepackage{jheppub} 

\usepackage{bbm, enumerate, framed}

\newcommand{\power}{\lambda}
\newcommand{\propconst}{\xi}

\preprint{LTH 1016, \ ZMP-HH/14-16}

\title{\boldmath Non-extremal black hole solutions from the $c$-map}

\author[a]{D. Errington,}
\author[a]{T. Mohaupt,}
\author[b]{O. Vaughan}

\affiliation[a]{University of Liverpool,\\Department of Mathematical Sciences, Peach Street, Liverpool L69 7ZL, UK}
\affiliation[b]{University of Hamburg,\\Department of Mathematics, Bundesstra{\ss}e 55, D-20146 Hamburg, Germany}

\emailAdd{david.errington@liv.ac.uk}
\emailAdd{thomas.mohaupt@liv.ac.uk}
\emailAdd{owen.vaughan@math.uni-hamburg.de}

\abstract{We construct new static, spherically symmetric non-extremal black hole 
solutions of four-dimensional ${\cal N}=2$ supergravity, using a systematic technique based on dimensional reduction over time (the $c$-map) and the real formulation of special geometry.
For a certain class of models we actually obtain the general solution to the full second order equations of motion, whilst for other classes of models, such as those obtainable by dimensional reduction from five dimensions, 
heterotic tree-level models, and type-II Calabi-Yau compactifications
in the large volume limit a partial set of solutions are found.
When considering specifically non-extremal black hole solutions we find that regularity conditions reduce the number of integration constants by one half.
Such solutions satisfy a unique set of first order equations, which we identify. 

Several models are investigated in detail, including examples of non-homogeneous spaces such as the quantum deformed $STU$ model.
Though we focus on static, spherically symmetric solutions
of ungauged supergravity, the method is 
adaptable to other types of solutions and to gauged supergravity.}

\begin{document}

\vskip 1.5 true cm  

\maketitle
\flushbottom

\section{Introduction}

Understanding non-extremal black holes
in terms of string theory is the next major step after the
earlier discovery of the deep relation between BPS black holes solutions 
and BPS excitations of strings and branes.
Most of the current literature focuses on two ideas: (i) concentrating on models where the scalar manifold is a symmetric space and generating the general solution using group
theoretical methods \cite{Bergshoeff:2008be,Chemissany:2009hq,Chemissany:2010zp}\footnote{Note that in $N=4,8$ supergravity the target manifold is always symmetric. For string theory compactifications with these symmetries the most general BPS and non-extremal black hole solutions have been known for some time \cite{Cvetic:1995uj, Cvetic:1995bj, Cvetic:1996xz, Cvetic:1996kv}.}; (ii) reducing the full
second order scalar equations of motion to first order gradient flow
equations, thus obtaining a structure similar to 
BPS solutions \cite{Lu:2003iv,Miller:2006ay,Ceresole:2007wx,Andrianopoli:2007gt,Janssen:2007rc,Cardoso:2008gm, Perz:2008kh,Gutowski:2008tm,Barisch:2011ui,Goldstein:2014qha}.

In this paper we continue developing a complementary approach which 
for five-dimensional solutions was developed in 
\cite{Mohaupt:2009iq, Mohaupt:2010fk, Mohaupt:2012tu,Dempster:2013mva},
and for four-dimensional solutions in 
\cite{Mohaupt:2011aa}. We do 
not assume that the scalar target space is a Riemannian symmetric space,
nor that it is homogeneous, but work in the framework of special geometry,
which applies to any ${\cal N}=2$ supergravity theory and
string compactification. Moreover we directly solve the second order
field equations and for the subclass of so-called diagonal 
models (which includes the $STU$-model, along with other models
with non-homogeneous target spaces) we even obtain the
most general spherically symmetric solution with purely imaginary 
scalar fields and half of the gauge charges turned on. 
We then observe that solutions which correspond
to black holes subject to suitable regularity conditions
depend only on half of the number 
of possible integration constants, and  satisfy a unique set of first
order equations, which we identify. 
This re-enforces
the view that non-extremal solutions preserve some of the
features known from BPS solutions. 

The fact that certain non-extremal black hole solutions obey unique first order equations has been known for some time in the literature, e.g.\ \cite{Lu:2003iv, Miller:2006ay, Janssen:2007rc, Andrianopoli:2007gt}, but the first order rewriting
is imposed as an ansatz and does not exclude the existence of more general
non-extremal solutions which cannot obtained this way. 
In our approach the logic is different: we first find a solution to the full second order equations of motion, 
and then restrict these solutions to those that correspond to non-extremal black holes. We find that these solutions \emph{must} satisfy a unique set of first order equations. Moreover, since we observe this feature for a large class of models we expect this to be a common feature for all static, spherically symmetric non-extremal black hole solutions in ${\cal N} = 2$ supergravity coupled to vector multiplets.

Let us explain the key concepts of our approach, which have been discussed in detail in \cite{Mohaupt:2011aa}. Since we are interested
in stationary solutions, we perform a dimensional reduction over time and
work with the resulting effective three-dimensional Euclidean theory. 
Three-dimensional gravity has no local dynamics, and we can dualise
three-dimensional abelian gauge fields into scalars, which leaves us 
with a non-linear sigma model with some target space $\bar{N}$.
Starting with four-dimensional ${\cal N}=2$ vector multiplets,
where the target space is a projective special 
K\"ahler (PSK) manifold $\bar{M}$, the space $\bar{N}$ is a para-quaternionic 
K\"ahler manifold \cite{Cortes:2003zd,Vaughan:2012,Cortes:2014}, and the relation between $\bar{M}$ and $\bar{N}$ is the
temporal version of the $c$-map. Restricting our attention to
spherically symmetric solutions, solving the four-dimensional 
equations of motion reduces to the problem of finding harmonic
maps from the reduced three-dimensional space-`time' to the manifold
$\bar{N}$, the image being a geodesic curve parametrised by the three-dimensional
scalar fields \cite{Breitenlohner:1987dg,Cortes:2009cs}.\footnote{In gauged
supergravity the geodesic equation is modified by a potential, the 
inclusion of which into the formalism was discussed in \cite{Klemm:2012yg, Klemm:2012vm}.} 
We refer to solutions of the 
three-dimensional Euclidean theory as instantons, although we do
not verify explicitly that they have a finite action. 
Upon lifting these solutions to four dimensions, we find that a
subset correspond to regular black hole solutions. Based on 
the results of \cite{Cortes:2009cs,Mohaupt:2010du} we expect that 
after adding a suitable
boundary term at least this subset of solutions
will have a finite Euclidean action which is related to the
ADM mass of the black hole. We refer to extremal (non-extremal)
instantons as solutions 
which lift to extremal (non-extremal) black holes. These solutions
correspond to null
(non-null) curves in the scalar manifold $\bar{N}$.

We find it useful to use the real formulation of special geometry
developed in \cite{Mohaupt:2011aa}, which is based on a real
Hesse potential $H$, rather than the more familiar
formulation based on a holomorphic prepotential $F$. The real formulation
leads to a more transparent parametrisation of the manifold $\bar{N}$, which
in particular allows one to preserve symplectic covariance. 
In this paper we extend the results of \cite{Mohaupt:2011aa} to
non-extremal solutions
by identifying conditions that lead to an explicit calculation of the Hesse potential and a 
simplification of the equations of motion. Specifically we will impose that field configurations are spherically symmetric and that the four-dimensional complex scalar fields, typically denoted $z^A$, are purely
imaginary, or PI, a condition which for models obtainable by reduction from five dimensions is known as axion-free or non-axionic. 
The PI conditions freeze half of the scalars and eliminate half of the
charges. This simplifies the equations of motion to the extent that 
they take a form similar to the five dimensional case considered 
before in \cite{Mohaupt:2009iq, Mohaupt:2010fk,Dempster:2013mva}.
Specifically, after imposing spherical symmetry and the PI conditions, 
the equations of motion can be obtained from variation of the 
one-dimensional effective Lagrangian
\[
	{\cal L} = \tilde{H}_{ab}(q)\left( \dot{q}^a \dot{q}^b - \dot{\hat{q}}^a \dot{\hat{q}}^b \right) \;,
\]
together with imposing the Hamiltonian constraint. Here
the three-dimensional scalar fields $(q^a, \hat{q}^a)$ parametrise
a pseudo-Rie\-mann\-ian manifold equipped with a Hessian metric 
$\tilde{H}_{ab}$.

In order to solve the corresponding equations of motion we
observe that they decouple into self-contained subsets whenever
the scalar metric $\tilde{H}_{ab}$ exhibits a block structure. 
For each irreducible block we can find at least one independent solution, which contains two free integration constants, in
closed form. Thus if the scalar metric $\tilde{H}_{ab}$ decomposes into $m>1$ blocks,
we can find a solution which depends on $m$ independent three-dimensional 
scalar fields. The solution still depends on all $n + 1$ charges allowed by the
PI conditions, but the ratios between scalar fields belonging to the same
block are determined by the ratios of the corresponding gauge charges.

Throughout this paper we will focus on models with prepotentials of the form 
\begin{equation}
\label{eq:Ff1}
	F = i^{\power-1} \frac{f(Y^1,\ldots, Y^n)}{(Y^0)^\power} \;,  
\qquad \power,\,n \in {\mathbbm Z}^{>0} 
\end{equation}
where the holomorphic function $f$ is such that it is real-valued
when evaluated on real fields $Y^I$. Since $F$ is required to be homogeneous of degree two, $f$ must be homogeneous of degree $\power +2$.
For this class of models the scalar metric decomposes into at least two blocks, so that we may construct three-dimensional solutions with at least two independent scalar fields, which lift to four-dimensional solutions with at least one non-constant scalar field. While this is not the most general type of solution, it still represents an interesting new type of solution to a large class of models. In particular, models obtainable by dimensional reduction 
from five dimensions have prepotentials of the form (\ref{eq:Ff1}) with
$\power =1$ and $f=c_{ABC} Y^A Y^B Y^C$. 
By means of the M-theory limit and mirror symmetry, any type-II 
Calabi-Yau compactification takes this form asymptotically
in the large volume/large complex structure limit. 

As a concrete example we will consider  the quantum deformed $STU$ model with prepotential $F =- \frac{Y^1 Y^2 Y^3 + a(Y^3)^3}{Y^0}$.
This can be realised, with $a=\frac{1}{3}$, as a heterotic string 
compactification on $K3\times T^2$ with instanton numbers 
$(12,12)$ or $(13,11)$ or $(14,10)$. In this realisation the
term proportional to $(Y^3)^3/Y^0$ arises as a one loop correction
\cite{deWit:1995zg,Harvey:1995fq}. 
Equivalently, the same model can be obtained 
as a type-IIA compactification 
on a certain family of elliptically fibred Calabi-Yau three-folds
with basis the Hirzebruch surfaces $\mathbbm{F}_0$ or $\mathbbm{F}_1$
or $\mathbbm{F}_2$ \cite{Louis:1996mt}. 
In this case all contributions to the prepotential arise at the
classical level. 
We will use this model frequently as an example, as it provides a 
simple non-homogeneous deformation of the symmetric $STU$-model, and 
is one of the simplest examples to study the heterotic/type-II string 
duality.

Whilst we may generically construct solutions to models of the form 
(\ref{eq:Ff1}) with two independent scalar fields, more general solutions are possible if the scalar metric decomposes into more than two blocks. 
One interesting and relevant class are prepotentials
which are linear in one field, say $Y^1/Y^0$, and thus have the structure
\begin{equation}
F = \frac{f_1(Y^1) f_2(Y^2, \ldots Y^n)}{Y^0} \;, 
\label{HetPrepo}
\end{equation}
where $f_1$ and $f_2$ have degree one and two. This class includes
all tree-level heterotic prepotentials, which are always linear
in the dilaton (see for example \cite{deWit:1995zg}),
${\cal N}=2$ truncations of ${\cal N}=4$ supergravity (see for
example \cite{Mohaupt:2000mj}), and 
models based on  reducible Jordan algebras (see for example \cite{Gutowski:2008tm}). 
For prepotentials of the form (\ref{HetPrepo}) we will show that
the scalar metric decomposes into three independent
blocks, so that we can obtain three-dimensional 
non-extremal solutions with three independent
scalars.

The limiting case is given by `diagonal' models, in which the scalar equations of motion decouple completely from one-another and it is possible to obtain the most general stationary solution that satisfies the spherically symmetric and PI conditions. Diagonal models are characterised by prepotentials of the 
form\footnote{One can of course obtain equivalent formulations of these
models by applying symplectic transformations.}
\begin{equation}
F = i^{\power-1} \frac{(Y^1\ldots Y^n)^{\frac{\power + 2}{n}}}{(Y^0)^\power} \;,
\qquad \power,\,n \in {\mathbbm Z}^{>0} 
\;. \notag 
\end{equation}
It is known from \cite{deWit:1991nm} that homogeneous special K\"ahler
spaces either have prepotentials of the very special 
form $F=\frac{c_{ABC}Y^A Y^B Y^C}{Y^0}$ or are in the ${\mathbbm C}H^n$ series,
where the prepotential is not of diagonal type. It follows that within
the diagonal class there 
are precisely two homogeneous spaces, given by 
\begin{equation*}
	F = \frac{(Y^1)^3}{Y^0}\;, \qquad
	F = \frac{Y^1 Y^2 Y^3}{Y^0}\;, 
\end{equation*}
which correspond to the symmetric spaces  $SU(1,1)/U(1)$ and $[SU(1,1)/U(1)]^3$ respectively. All other diagonal models are therefore not homogeneous.
As a concrete example of a non-homogeneous diagonal model 
we will consider the prepotential $F = i\frac{Y^1 Y^2 Y^3 Y^4}{(Y^0)^2}$.

We will also discuss one solution that is valid for generic models, i.e.\ for any choice of holomorphic prepotential $F$, which we will refer to as the universal solution. In this case the three-dimensional scalar fields are all proportional to one-another. In four-dimensions this solutions is characterised by a Reissner--Nordstr\"om spacetime metric, $n + 1$ electric and magnetic charges, and constant four-dimensional scalar fields $z^A$.


The formalism used in this paper has been adapted to 
gauged supergravity \cite{Klemm:2012yg, Klemm:2012vm} and to the construction of non-extremal rotating solutions \cite{Gnecchi:2013mja}.
There are similarities between our approach and the 
H-FGK approach of \cite{Meessen:2011aa}, which builds
on \cite{Ferrara:1997tw}. In particular both
methods use adapted variables to preserve symmetries and do not rely 
on group theoretic methods.
The H-FGK method was recently
used to obtain solutions to type-II models with $\alpha'$-corrections
\cite{Galli:2012pt, Bueno:2012jc}.

This paper is organised as follows: in Section \ref{VM+c}
we provide the necessary background on vector multiplets and 
dimensional reduction (the $c$-map). Key results on Hessian metrics,
some of which are not available in the existing literature, 
are collected in Appendix \ref{App:Hessian}. In Section 
\ref{SphericalPI} we analyse field configurations which are
spherically symmetric with four-dimensional scalars restricted to
purely imaginary values, and derive the resulting simplifications
of the three-dimensional equations of motion. Some auxiliary results
on spherically symmetric metrics are reviewed in 
Appendix \ref{SphericalSym}. In Section \ref{sec:3dsols} we solve
the three-dimensional equations of motion, while in Section \ref{sec:lifting}
we lift these solutions to four dimensions and determine which of these 
correspond to black holes. While we solve the full second order equations
of motion we demonstrate in Section \ref{sec:BHIntConsts} that after 
imposing the regularity conditions required to obtain four-dimensional
black holes, our solutions satisfy first order equations. Our conclusions
are presented in Section \ref{ConOut}.

Throughout this paper $n$ will denote the number of vector multiplets and our index conventions will be:

\begin{center}
    \begin{tabular}{ | l | l | l | p{5cm} |}
    \hline
    Spacetime indices &
		\hspace{1.95em}$\hat{\mu}, \hat{\nu}, \ldots = 0,1,2,3$ \\
		&	\hspace{1.7em} $\mu, \nu, \ldots = 1,2,3$ \\
		%
		\hline 
		Target space indices & 
		  \hspace{0.5em} $A,B,C\ldots = 1, \ldots, n$ \\
		& \hspace{0.65em} $I,J,K,\ldots = 0, \ldots, n$ \\
		& \hspace{1.2em} $a,b,c, \ldots = 0, \ldots, 2n + 1$ \\
		& \hspace{1.95em} $\rho, \sigma, \ldots = 1, \ldots, n + 1$ \\
		& \hspace{0.75em} $\alpha, \beta, \gamma, \ldots = 0, n + 2, \ldots, 2n + 1$ \\
		& \hspace{0.0em} $\alpha', \beta', \gamma', \ldots = n + 2, \ldots, 2n + 1$ \\
		\hline
    \end{tabular}
\end{center}

\section{Review of vector multiplets and of the $c$-map \label{VM+c}}

In this section we review the special geometry of four-dimensional
vector multiplets and their dimensional reduction over time (the
temporal version of the $c$-map.) This is mostly based on 
\cite{Mohaupt:2011aa}, though we also derive new explicit expressions
for the inverse Hessian metrics $H^{ab}$ and $\tilde{H}^{ab}$ in 
(\ref{Hinvab}), (\ref{Htildeinv2a}), (\ref{tildeHabformula1a}),
which are proved in Appendix \ref{App:Hessian}.

\subsection{Four-dimensional vector multiplets}\label{VMsec}

The couplings of four-dimensional ${\cal N}=2$ vector multiplets
to supergravity were constructed in \cite{deWit:1984pk} using the 
conformal calculus. We refer to \cite{Freedman:2012zz} for a detailed
review. The approach to special geometry taken in \cite{Mohaupt:2011aa}
is based on the conformal calculus, combined with more recent work
in differential geometry, in particular \cite{Freed:1997dp}, 
\cite{Alekseevsky:1999ts}.

The bosonic part of the four-dimensional vector multiplet Lagrangian is%
\footnote{The sign of the Einstein-Hilbert term is different to \cite{Mohaupt:2011aa}, due to the fact that in this paper we define the Riemann tensor by $R^{\rho}_{\;\;\lambda \mu \nu} = \partial_\mu \Gamma^\rho_{\;\;\nu \lambda} - \partial_\nu \Gamma^\rho_{\;\;\mu \lambda}
		+ \Gamma^\rho_{\;\;\mu \sigma} \Gamma^\sigma_{\;\;\nu \lambda} -  \Gamma^\rho_{\;\;\nu\sigma} \Gamma^\sigma_{\;\;\mu\lambda}\,.$}
%
\begin{equation}
		\texttt{e}_4^{-1}{\cal L}_4 = \tfrac{1}{2} R_4 
- g_{A\bar{B}} \partial_{\hat{\mu}} z^A \partial^{\hat{\mu}} \bar{z}{}^{\bar{B}} 
+ \tfrac{1}{4} {F}^I_{\hat{\mu} \hat{\nu}} \tilde{G}_{I|\hat{\mu}\hat{\nu}} \;, 
\label{eq:4dSugraLag}
\end{equation}
where $R_4$ and $\texttt{e}_4$ are the four-dimensional Ricci scalar and
vielbein, $\hat{\mu}, \ldots = 0,\ldots 3$ are 
four-dimensional space-time indices, 
$z^A$ with $A=1, \ldots, n$ are complex scalars, $F^I_{\hat{\mu}\hat{\nu}}$ with $I=0,\ldots, n$ are abelian field strengths
and 
\[
	G_{I|\hat{\mu}\hat{\nu}} := {\cal R}_{IJ}  {F}^J_{\hat{\mu} \hat{\nu}} - {\cal I}_{IJ} \tilde{{F}}^J_{\hat{\mu} \hat{\nu}} \;.
\]
The scalar fields $z^A$ parametrise
a projective special K\"ahler (PSK) manifold $\bar{M}$. All couplings
in the Lagrangian are encoded in the holomorphic prepotential 
${\cal F}(z^A)$. 
The scalar couplings are given by the metric of the PSK manifold $\bar{M}$:
\[
g_{A\bar{B}} = \partial_A \partial_{\bar{B}} K_{\bar{M}} \;,\;\;\;
K_{\bar{M}} = - \log \left( 2i ({\cal F}-\bar{\cal F}) -
i (z^A -\bar{z}^{\bar{B}}) ({\cal F}_A + \bar{\cal F}_{\bar{B}})\right)\;,
\]
where we use the notation 
${\cal F}_A = \partial_A {\cal F}$, etc.
The vector field couplings are encoded in the complex matrix
${\cal N}_{IJ}= {\cal R}_{IJ} + i {\cal I}_{IJ}$, which is defined below
in (\ref{VectorCouplings}).

The field equations, though not the Lagrangian itself, are invariant
under symplectic $Sp(2n+2,\mathbbm{R})$ transformations, which 
generalise the electric-magnetic duality of Maxwell theory. This
becomes more transparent when using the gauge-equivalent description 
of the theory in terms of $n+1$ superconformal vector multiplets. 
Denoting the superconformal scalars as $X^I$, $I=0,\ldots, n$, the
couplings of such a theory are encoded in a holomorphic 
prepotential $F(X^I)$, which is,
in addition, homogeneous of degree two:
$F(\lambda X^I) = \lambda^2 F(X^I)$, where $\lambda \in \mathbbm{C}^*$. 
The associated scalar metric is 
\[
N_{IJ} = 2 \mbox{Im}F_{IJ} = \frac{\partial K_M}{\partial X^I \partial
\bar{X}^J} \;,
\]
with K\"ahler potential
\[
K_M = i (X^I \bar{F}_I - F_I \bar{X}^I)\;.
\]
Here we use a notation where $F_I = \frac{\partial F}{\partial X^I}$, etc.
The scalars $X^I$ parametrise a conical affine special K\"ahler (CASK)
manifold $M$, which is a complex cone over $\bar{M}$.
The vector couplings in general involve $R_{IJ}=2 \mbox{Re} F_{IJ}$ as 
well as $N_{IJ}$. 

Symplectic transformations act by matrices ${\cal O} = ({\cal O}^a_{\;b}) 
\in Sp(2n+2,\mathbbm{R})$,  which are defined by
\[
{\cal O}^T \Omega {\cal O} = \Omega \;,
\]
where 
\[
\Omega=
\left( \begin{array}{cc}
0 & \mathbbm{1} \\
-\mathbbm{1} & 0 \\
\end{array} \right) \;.
\]
In the complex formulation of special geometry, the quantities
$(X^I,F_I)^T$ and $(F^I_{\hat{\mu} \hat{\nu}}, G_{I|\hat{\mu} \hat{\nu}})^T$
transform as vectors, while 
$i (X^I \bar{F}_I - F_I \bar{X}^I)$ transforms as a scalar (function).
However the holomorphic prepotential $F$ does not transform as a scalar,
and  $N_{IJ}$ and ${\cal N}_{IJ}$ transform by fractional linear 
transformations. We will therefore later introduce the real formulation
of special geometry where all relevant quantities transform as tensors.

A gauge equivalent formulation of 
(\ref{eq:4dSugraLag}) is obtained by gauging the superconformal 
symmetries.  After eliminating various
auxiliary fields by their equations of motion, the 
bosonic part of the superconformal Lagrangian is
\begin{equation}
\label{4dscLagr}
\texttt{e}_4^{-1}{\cal L}_4 = \tfrac{1}{2} e^{-{\cal K}(X)} R_4 
- e^{-{\cal K}(X)} g_{IJ} \partial_{\hat{\mu}} X^I \partial^{\hat{\mu}}
\bar{X}^J + \frac{1}{4} e^{-{\cal K}(X)}\partial_{\hat{\mu}} {\cal K}
\partial^{\hat{\mu}} {\cal K}
+ \tfrac{1}{4} {F}^I_{\hat{\mu} \hat{\nu}} \tilde{G}_{I|\hat{\mu}\hat{\nu}} \;, 
\end{equation}
where
\begin{equation}\label{eq:curlyKX}
e^{-{\cal K}(X)} = - i (X^I \bar{F}_I - F_I \bar{X}^I) \;.
\end{equation}
The scalar couplings are 
\[
g_{IJ} = 
\frac{\partial^2 {\cal K}}{\partial X^I \partial \bar{X}^J} =
\frac{N_{IJ}}{(-X^M N_{MN} \bar{X}^N)} + 
\frac{N_{IK} \bar{X}^K N_{JL} {X}^L}{(-X^M N_{MN} \bar{X}^N)^2} \;,
\]
while the vector couplings are
\begin{equation}
\label{VectorCouplings}
{\cal N}_{IJ} = {\cal R}_{IJ} + i {\cal I}_{IJ} =
\bar{F}_{IJ} + i \frac{N_{IK} X^K N_{JL} X^L}{N_{MN} X^M X^N} \;.
\end{equation}

The Lagrangians (\ref{4dscLagr}) and (\ref{eq:4dSugraLag})
are gauge equivalent due to the $\mathbbm{C}^*$-trans\-for\-ma\-tions
acting on (\ref{4dscLagr}). The infinitesimal generators of the
$\mathbbm{C}^*$-action are the vector fields
\[
\xi = X^I \frac{\partial}{\partial X^I} + \bar{X}^I 
\frac{\partial}{\partial \bar{X}^I} \;,\;\;\;
J\xi = i X^I \frac{\partial}{\partial X^I} - i \bar{X}^I 
\frac{\partial}{\partial \bar{X}^I} \;,
\]
where $J$ denotes the complex structure of $M$. The resulting 
finite transformations are
\[
X^I \mapsto \lambda X^I = |\lambda| e^{i\theta} X^I \;,\;\;\;
\lambda \in \mathbbm{C}^* \;.
\]
The real scale transformations generated by $\xi$ are homotheties
of the CASK metric $N_{IJ}$ while the $U(1)$ transformations
generated by $J\xi$ are isometries.

To recover (\ref{eq:4dSugraLag}) from (\ref{4dscLagr})
one needs to gauge-fix these transformations.
The first step is to impose the D-gauge 
\[
e^{-{\cal K}(X)}= - i (X^I \bar{F}_I - F_I \bar{X}^I)=1 \;,
\]
which fixes the real scale transformations $\mathbbm{R}^{>0}\subset
\mathbbm{C}^*$ 
and brings the Einstein-Hilbert 
term to its canonical form. The second step is to fix the remaining
$U(1) \subset \mathbbm{C}^*$ transformations. Fixing a $U(1)$ 
gauge necessarily requires giving up manifest symplectic covariance.
Therefore we postpone this step and formulate the $c$-map in 
a formalism with manifest symplectic and $U(1)$ covariance. Later, when
we construct solutions, the restriction to purely imaginary field 
configurations will force us to fix a $U(1)$ gauge, which will 
be done by imposing $\mbox{Im}X^0=0$. This, and as well any further
condition we impose on solutions, restricts symplectic covariance
to the subgroup commuting with all conditions.

The gauge equivalence between (\ref{eq:4dSugraLag}) and
(\ref{4dscLagr}) implies that after imposing the D-gauge
and taking into account the residual $U(1)$ symmetry 
the fields $X^I$ only represent $2n$ rather than $2n+2$ 
independent real degrees of freedom. This is seen by observing
that the tensor $g_{IJ}$ is $\mathbbm{C}^*$-invariant and 
has a two dimensional kernel 
(since $X^I g_{IJ} = 0 = g_{IJ} \bar{X}^J$), which makes modes
corresponding to $\mathbbm{C}^*$-transformations non-propagating. 
The standard way of obtaining (\ref{eq:4dSugraLag}) from
(\ref{4dscLagr}) is to introduce inhomogeneous special coordinates
$z^A = X^A/X^0$, which are $\mathbbm{C}^*$-invariant, and to verify
that (in the D-gauge) $g_{IJ} \partial_{\hat{\mu}} X^I \partial^{\hat{\mu}}
\bar{X}^J = g_{A\bar{B}} \partial_{\hat{\mu}} z^A \partial^{\hat{\mu}} 
\bar{z}^{\bar{B}}$. Geometrically, 
the degenerate tensor $g_{IJ}$ is the horizontal
lift of the PSK metric $g_{A\bar{B}}$
on $\bar{M}$ to the complex cone $M$.
The prepotentials $F(X^I)$ and ${\cal F}(z^A)$
are simply related by ${\cal F}(z^A) = (X^{0})^{-2} F(X^I) =
F(1,z^A)$.
For the vector kinetic terms one uses that 
${\cal N}_{IJ}$ is homogeneous of degree zero, and therefore 
${\cal N}_{IJ} (X^K) = {\cal N}_{IJ}(1, z^A)$.

\subsection{Special real coordinates}

We now review and extend the formulation of special K\"ahler geometry in terms
of special real coordinates given in \cite{Mohaupt:2011aa}. Special
real coordinates were introduced in 
\cite{Freed:1997dp,1999math......1069H,Alekseevsky:1999ts}, and later
used in work on black hole solutions and higher derivative corrections
\cite{LopesCardoso:2006bg,Cardoso:2010gc,Cardoso:2012nh,Cardoso:2012mc,Cardoso:2014kwa}. The formalism of \cite{Mohaupt:2011aa} provides a formulation
of special K\"ahler geometry in terms of special real coordinates
on the CASK manifold
associated with the gauge-equivalent superconformal theory. This has
the advantage to fully preserve symplectic covariance and can be
viewed as an off-shell generalisation of the symplectically covariant
formalism used in \cite{Behrndt:1996jn,LopesCardoso:1998wt,LopesCardoso:2000qm}
to construct BPS black holes. A different real formalism, which uses
special real coordinates on the PSK manifold itself, was developed
in \cite{Ferrara:2006at}.

Special real coordinates on the CASK manifold $M$ are 
defined by
\begin{equation}
\label{SpecialReal1}
(q^a) := \left( \begin{array}{cc}
x^I \\ y_I \end{array} \right)
:= \mbox{Re} \left( \begin{array}{cc}
X^I \\ F_I \end{array} \right) \;,
\end{equation}
where $a=0,\ldots 2n+1$. In the real formulation all couplings are
encoded in the Hesse potential $H(q^a)$, which is related to the
prepotential $F(X^I)$ by a Legendre transformation
$(x^I, u^I) := (\mbox{Re}X^I, \mbox{Im} X^I) \rightarrow
(x^I, y_I)$:
\[
H(x^I, y_I) = 2 \mbox{Im} F(X^I(x,y)) - 2 \mbox{Im} F_I(x,y) y^I \;.
\]
Derivatives of the Hesse potential will be denoted 
$H_a = \frac{\partial H}{\partial q^a}$, etc. 
The Hessian metric 
\[
H_{ab} = \frac{\partial^2 H}{\partial q^a \partial q^b}
\]
is the real version of the K\"ahler metric 
$N_{IJ} = 2 \mbox{Im} F_{IJ}$ on $M$ in the
sense that $N_{IJ} dX^I d\bar{X}^J = H_{ab} dq^a dq^b$. 
In special real coordinates, the associated K\"ahler form
is simply
\[
\omega = 2 dx^I \wedge y_I = \Omega_{ab} dq^a \wedge dq^b \;.
\]
We will denote the inverse of this matrix by $\Omega^{-1} = (\Omega^{ab})$.
The special real coordinates are Darboux coordinates. The
complex structure takes the form\footnote{This is the standard relation
between the complex structure, the K\"ahler form and the metric of a K\"ahler
manifold. The factor $\frac{1}{2}$ is due to the fact that the matrix
representing the K\"ahler form $\omega(\cdot,\cdot) = g(\cdot,J\cdot)$ with respect to the coordinates
$q^a$ is $2\Omega_{ab}$.}
\[
J^a_{\;\;c} = -\frac{1}{2} \Omega^{ab} H_{bc} \;.
\]

In special real coordinates the infinitesimal action of $\mathbbm{C}^*=\mathbbm{R}^{>0}\cdot U(1)$ is generated by the vector fields\footnote{The special holomorphic coordinates $X^I$ 
and the associated special real coordinates $q^a$ are adapted to the
$\mathbbm{C}^*$-action, they are `conical special coordinates' 
\cite{Cortes:2009cs}. In terms
of such coordinates, which are unique up to linear symplectic
transformations, the prepotential and the Hesse potential are 
homogeneous of degree two with respect to complex and real scale
transformations, respectively.}
\[
\xi = q^a \frac{\partial}{\partial q^a} \;,\;\;\;J\xi = \frac{1}{2} 
H_a \Omega^{ab} \frac{\partial}{\partial q^b} \;.
\]
The Hesse potential is homogeneous of degree two under the real
scale transformations generated by $\xi$ and invariant under the
$U(1)$ transformations generated by $J\xi$.

Since
\[
e^{-{\cal K}(X)} = -i (X^I \bar{F}_I- F_I \bar{X}^I) = - 2H \;,
\]
the D-gauge corresponds to $-2H=1$. 

The finite transformations generated by $\xi$ and $J\xi$, respectively, 
are
\[
q^a \mapsto |\lambda| q^a \;,\;\;\;
q^a \mapsto \cos \theta \, q^a + \sin \theta (Jq)^a\;,
\]
where
\begin{equation}
\label{Jqa}
(Jq)^a = J^a_{\;\;b}q^b = -\frac{1}{2} \Omega^{ac} H_{cb} q^b\;.
\end{equation}

The Hessian metric $H_{ab}$ can be decomposed as
\[
H_{ab} = (-2H) H^{(0)}_{ab} + \frac{1}{2H} H_a H_b + \frac{2}{H} 
\Omega_{ac} q^c \Omega_{bd} \Omega^d \;.
\]
The tensor $H^{(0)}_{ab}$ is 
the real version of the degenerate
tensor $g_{IJ}$, 
\[
g_{IJ} dX^I d\bar{X}^J = H^{(0)}_{ab} dq^a dq^b \;.
\]
In other words, the scalar term of
the bosonic Lagrangian can be rewritten as
\[
e^{-{\cal K}(X)} g_{IJ} \partial_{\hat{\mu}} X^I \partial^{\hat{\mu}} \bar{X}^J
= (-2H) H^{(0)}_{ab} \partial_{\hat{\mu}} q^a \partial^{\hat{\mu}} q^b \;.
\]

A second, `dual' set of special real coordinates is given by
\[
q'_a = H_a = \frac{\partial H}{\partial q^a} = \left( 
\begin{array}{c}
2 v_I \\ -2 u^I \\
\end{array} \right) =
\left( \begin{array}{c}
2 \mbox{Im} F_I \\
-2 \mbox{Im} X^I \\
\end{array} \right)
\;,
\]
where $u^I = \mbox{Im} X^I$ and $v_I =\mbox{Im} F_I$.
Since $H$ is homogeneous of degree two, 
the special coordinates and dual special coordinates 
are related by
\[
q'_a = H_{ab} q^b \Leftrightarrow q^q = H^{ab} q'_b\;,
\]
where $H^{ab}$ denotes the inverse of the Hessian metric $H_{ab}$. 
We thus have two expressions for the CASK metric on $M$:
\[
g = H_{ab} dq^a dq^b = H^{ab} dq'_a dq'_b \;.
\]
In Appendix \ref{App:Hessian} we show that the inverse 
metric $H^{ab}$ is a Hessian metric with respect to the dual coordinates:
\begin{equation}
\label{Hinvab}
H^{ab} = \frac{\partial^2 H'}{\partial q'_a \partial q'_b} \;,
\end{equation}
and that the corresponding Hesse potential is 
$H'(q'):=H(q(q'))$. For notational
simplicity we will often simply write $H(q')$ instead of the
accurate $H'(q')$.\footnote{Note that $H(q)$ is in general not
invariant under the diffeomorphism $q^a \rightarrow q'_a$, so
that is important to interpret $H(q')$ as $H(q(q'))$. }

Upper indices
$a,b, \ldots$ transform with symplectic matrices 
${\cal O} = ({\cal O}^a_{\;b})$, while lower indices
transform with the contragradient matrices 
${\cal O}^{T,-1} =:({\cal O}_a^{\;\;b})$. In particular
$q^a$ transforms as a vector, $q'_a$ as a co-vector
and $H_{ab}$ and $H^{ab}$ as second rank co-tensors and 
tensors, respectively, while the Hesse potential transforms
as a scalar. The raising and lowering of indices with the metric $H_{ab}$
is consistent with symplectic transformations. Moreover, the contraction 
of tensors with $\Omega_{ab}$ and its inverse is also consistent 
with symplectic covariance because $\Omega_{ab}$ intertwines between
the fundamental and contragradient representation of the 
symplectic group, i.e. if $q^a$ is a symplectic vector then 
$\Omega_{ab} q^b$ is a symplectic co-vector. 
For example, 
according to (\ref{Jqa}) 
the complex structure $J$ acts on $M$ by the diffeomorphism
\[
q^a \mapsto 
(Jq)^a =  -\frac{1}{2} \Omega^{ac} H_{cb} q^b = 
 -\frac{1}{2} \Omega^{ac} q'_c = 
-\left( \begin{array}{c}
u^I \\  v_I \\
\end{array} \right) = 
-\left( \begin{array}{c}
 \mbox{Im} X^I \\  \mbox{Im} F_I \\
\end{array} \right) \;.
\]
Thus the vector $(Jq)^a$ and the co-vector 
$q'_a$ are related through multiplication by $\Omega_{ab}$.

\subsection{Dimensional reduction over time}
\label{sec:4dto3d}

The four-dimensional 
space-time metric $g^{(4)}$ and the three-dimensional Euclidean signature
metric $g^{(3)}$ are related by
\begin{equation}\label{eq:KKdecomp}
g^{(4)} = -e^\phi(dt + V_\mu dx^\mu)^2 + e^{-\phi} g^{(3)} \;,
\end{equation}
where $\phi$ is the Kaluza-Klein scalar and $V_\mu$ the 
Kaluza-Klein vector. It is useful to combine the Kaluza-Klein scalar
with the four-dimensional scalars, which in the superconformal formalism
are described by either the holomorphic fields $X^I$, or the real fields
$q^a$, 
subject to $\mathbbm{C}^*$-transformations. A key observation is that
$\phi$ can be identified with the radial degree of freedom of the cone
$M$ over $\bar{M}$, which thus is promoted from a gauge degree of
freedom to a physical degree of freedom \cite{Mohaupt:2011aa}. 
In terms of the holomorphic 
formulation, this is done by defining the rescaled complex symplectic
vector
\begin{equation}\label{eq:YtoX}
\left( \begin{array}{c} Y^I \\ F_I(Y)  \end{array} \right) :=
e^{\phi/2} \left( \begin{array}{c} X^I \\ F_I (X) \end{array} \right) \;.
\end{equation}
Here we used that $F_I=F_I(X)$ is homogeneous of degree one. In the following
we will mostly use the rescaled variables $Y^I$ and usually denote $F_I(Y)$ 
by $F_I$.

If we impose the D-gauge $-i(X^I \bar{F}_I - F_I \bar{X}^I) = 1$,
this implies that $-i(Y^I \bar{F}_I(Y) - F_I(Y) \bar{Y}^I) = e^{\phi}$,
which determines the Kaluza-Klein scalar in terms of the $Y^I$. As long
as we do not impose a $U(1)$ gauge the $Y^I$ are still subject to
$U(1)$ transformations, but the expression for $e^\phi$, and as well
the Lagrangian displayed below, are $U(1)$ invariant. The same rescaling
can be performed with special real coordinates, or, equivalently, 
we can modify the definition of special real
coordinates by decomposing the complex vector $(Y^I,F_I)^T$ rather than
$(X^I,F_I)^T$
\begin{equation}\label{eq:YF(Y)decomp}
\left( \begin{array}{c} x^I + iu^I \\ y_I + iv_I \end{array} \right) :=
\left( \begin{array}{c} Y^I \\ F_I(Y)  \end{array} \right) \;,
\end{equation}
\begin{equation}
\label{SpecialReal2}
(q^a) := \left( \begin{array}{cc}
x^I \\ y_I \end{array} \right)
:= \mbox{Re} \left( \begin{array}{cc}
Y^I \\ F_I(Y) \end{array} \right) \;.
\end{equation}
Note that from now on we use special real coordinates which are defined
by (\ref{SpecialReal2}) rather than (\ref{SpecialReal1}). Due to 
the homogeneity of the prepotential all formulas derived using 
(\ref{SpecialReal1}) are either preserved or modified in a way which is
completely determined by the scaling weights of the quantities involved.
In the real formalism the Kaluza-Klein scalar is given by 
\begin{equation}\label{ephi2H}
e^\phi =-2H = -i \left( Y^I \bar{F}_I(Y) - F_I(Y) \bar{Y}^I \right) \;.
\end{equation}

The three-dimensional theory further contains
the scalar fields 
\[\hat{q}^a
= \left(\frac{1}{2}\zeta^I,\frac{1}{2}\tilde{\zeta}_I\right)
\] 
which 
descend from the gauge field degrees of freedom. The relation between 
these three-dimensional scalars and the four-dimensional gauge fields
can most easily be described via their derivatives 
	\begin{align}
		\left( \begin{array}{c} \partial_\mu \zeta^I \\ \partial_\mu \tilde{\zeta}_I \end{array} \right) = 
		\left( \begin{array}{c} F_{\mu0}^{I} \\ G_{I|\mu 0} \end{array} \right)\;.
	\end{align}
While the scalars $\zeta^I$ correspond to the time-like components of
the four-dimensional vector fields $A^I_{\mu}$, the scalars 
$\tilde{\zeta}_I$ are obtained by dualising the reduced, three-dimensional
vector fields. To obtain a formulation where all propagating bosonic
degrees of freedom are scalars, we also dualise the KK-vector $V_\mu$
into a scalar field $\tilde{\phi}$: 
\[
	\partial_{[\mu}V_{\nu]} = \frac{1}{2H^2} \varepsilon_{\mu \nu \rho} \left(\partial^\rho \tilde{\phi} +\tfrac{1}{2}(\zeta^I \partial^\rho \tilde{\zeta}_I - \tilde{\zeta}_I \partial^\rho \zeta^I)\right) \;.
\]

In \cite{Mohaupt:2011aa} it was shown that 
the Lagrangian of the three-dimensional theory can be arranged to take 
the form
\begin{eqnarray}\label{L3qupstairs}
\begin{aligned}
		\texttt{e}^{-1}_3 {\cal L}_3 = \;\;&\tfrac{1}{2} R_3 - \tilde{H}_{ab} \left(\partial_\mu q^a \partial^\mu q^b -  \partial_\mu \hat{q}^a \partial^\mu \hat{q}^b \right) \\ 
		&- \frac{1}{H^2} \left( q^a \Omega_{ab} \partial_\mu q^b \right)^2 +  \frac{2}{H^2} \left( q^a \Omega_{ab} \partial_\mu \hat{q}^b \right)^2  \\
		&- \frac{1}{4 H^2} \left( \partial_\mu \tilde{\phi} + 2\hat{q}^a \Omega_{ab} \partial_\mu \hat{q}^b  \right)^2 \;.
\end{aligned}
\label{eq:3dLag} 
\end{eqnarray}
Here $H$ is the Hesse potential, which depends on the rescaled special
real coordinates $q^a$ and encodes the KK-scalar, and 
\begin{equation}\label{tildeHformula}
{\tilde{H}_{ab}} := \frac{\partial^2}{\partial q^a \partial q^b} \tilde{H}
\;,\;\;\;
\tilde{H} := -\frac12 \log\left( -2H \right) \;.
\end{equation}
This tensor can be viewed as a modified metric on $M$, which
has been obtained by, essentially, replacing the Hesse potential 
by its logarithm. 
We remark that $\tilde{H}_{ab}$ is by construction a symplectic tensor,
and that raising and lowering tensor indices using $\tilde{H}_{ab}$ is
consistent with symplectic covariance. 

We will rely on various properties of the metric $\tilde{H}_{ab}$,
which are reviewed or derived in Appendix \ref{sec:dualcoords}. 
Here we only mention that it will be convenient later to use
dual coordinates with respect to $\tilde{H}_{ab}$ defined by
\[
q_a := \tilde{H}_a := \frac{\partial \tilde{H}}{\partial q^a} =
\frac{q'_a}{-2 H}  \;.
\]
where $q'_a=H_a$ are the dual coordinates with respect to $H$.  
Note that since $\tilde{H}_a$ is homogeneous of degree $-1$:
\begin{equation}
\tilde{H}_{ab} q^b = - \tilde{H}_a = - q_a \Rightarrow
q^a = - \tilde{H}^{ab}  q_b \;.
\end{equation}
One can show that $-\tilde{H}'(q_a) := - \tilde{H}(q^b(q_a))$
is a Hesse potential for the inverse metric $\tilde{H}^{ab}$,
\begin{equation}
\label{Htildeinv2a}
\tilde{H}^{ab} = \frac{\partial q^a}{\partial q_b}
= \frac{\partial^2 ( - \tilde{H}')}{\partial q_a \partial q_b}  \;.
\end{equation}
In practice we will compute $\tilde{H}^{ab}$ in terms of 
$H''(q_a):= H(q^b(q_a))$ by 
\begin{equation}
\tilde{H}^{ab} = -\frac{1}{2} 
\left( \frac{1}{H''} \frac{\partial^2 H''}{\partial q_a
\partial q_b} - \frac{1}{H^2} \frac{\partial H''}{\partial q_a}
\frac{\partial H''}{\partial q_b} \right) \;.
\label{tildeHabformula1a}
\end{equation}
For notational simplicity we will in the following write
$H(q_a)$ instead of $H''(q_a)=H(q^b(q_a))$ and $\tilde{H}(q_a)$
instead of $\tilde{H}(q^b(q_a))$. Note that in general neither
$H$ nor $\tilde{H}$ are invariant functions under the diffeomorphism
$q^a \mapsto q_a$.

The Lagrangian (\ref{eq:3dLag}) is invariant under 
symplectic transformations and local $U(1)$ transformations. It 
depends on $4n+5$ scalars $(q^a, \hat{q}^a, \tilde{\phi})$, but due
to the $U(1)$ gauge symmetry there are only $4n+4$ independent
propagating scalar degrees for freedom. One can gauge fix the
$U(1)$ symmetry by imposing any condition which is transversal
to the $U(1)$ action, and obtain a formulation in terms of 
$4n+4$  `physical' scalar fields. However, such a condition 
cannot be symplectically invariant and therefore breaks 
the manifest full symplectic covariance \cite{Mohaupt:2011aa}. 
Finding explicit solutions will require to gauge-fix the 
$U(1)$ at some point. In our case the gauge-fixing will be 
implied by a reality condition that we impose on solutions in order
to simplify the equations of motion. The solution will still be
expressed in terms of symplectic vectors, and manifest invariance
under the subgroup of symplectic transformations preserving the
reality condition will be preserved. This illustrates that while
any transversal condition can be used in principle to fix the $U(1)$,
the type of solution one wants to find typically selects a natural
gauge fixing condition. Thus one should not fix a $U(1)$ gauge too early.
Geometrically,
the $4n+5$ scalar fields are coordinates on the total space
of a $U(1)$ principal bundle $P$ over the $4n+4$-dimensional
scalar manifold $\bar{N}$ of the three-dimensional theory. Choosing
a $U(1)$ gauge allows one to embed $\bar{N}$ into $P$ as a submanifold.

\section{Purely imaginary 
and spherically symmetric field configurations 
\label{SphericalPI}}
 
In this section we will analyse the equations of motion given by the variation of (\ref{eq:3dLag}). The full field equations are given in 
Section 6.1 of \cite{Mohaupt:2011aa}, where a class of four-dimensional
stationary solutions were considered. In this paper we will impose two further
conditions which greatly simplify the equations of motion, namely that 
four-dimensional field configurations are 
\begin{enumerate}
	\item Purely imaginary,
	\item Spherically symmetric.
\end{enumerate}
The first is a condition on the target manifold, 
whilst the second is a condition on spacetime. Let us discuss each condition in turn and investigate the effect they have on the equations of motion.

\subsection{Purely imaginary field configurations}

We will call field configurations \emph{purely imaginary} 
if the complex PSK scalars $z^A$ are purely imaginary. Since $z^A = X^A/X^0=Y^A/Y^0$ and we choose the $U(1)$ gauge fixing condition $\mbox{Im} Y^0 = u^0 = 0$ as mentioned in Section \ref{VMsec}, this is equivalent to requiring that $Y^A$ are purely imaginary, or in other words 
\begin{equation}
	x^A = 0 \,,\;{A=1,\ldots,n} \;.
	\label{eq:PI}
\end{equation}
For models obtainable by dimensional reduction 
from five dimensions, the prepotential takes the very special form $F =\frac{c_{ABC}Y^A Y^B Y^C}{Y^0}$, with 
real $c_{ABC}$. In this case
the real parts of $z^A$ have an axion-like shift symmetry $z^A \mapsto z^A + \lambda^A$, and therefore purely imaginary configurations are sometimes referred to as \emph{axion-free} configurations. In this paper we will be interested in a more general class of models in which the prepotential takes the form
\begin{equation}
	F = i^{\power-1} \frac{f(Y^1,\ldots, Y^n)}{(Y^0)^\power} \;,
\label{eq:Ff}
\end{equation}
where $f$ is homogeneous of degree $\power+2$ and real when evaluated on real fields. For the particular choice $\power = 1$ and $f$ a cubic polynomial, this reduces to the class of models obtainable from five dimensions.

For models of the form (\ref{eq:Ff}) the purely imaginary (`PI') condition (\ref{eq:PI}) implies that $F_0$ is purely imaginary, or in other words $y_0 = 0$. Denoting by PI the restriction to purely imaginary configurations we have for this class of models
\begin{equation}
	(q^a)_{a = 0,\ldots,2n+1}\big|_{PI} = (x^0,0,\ldots,0;0,y_1,\ldots,y_n) \;,
	\label{eq:qPI}
\end{equation}
and by acting with the complex structure $J$ one finds 
\begin{equation}
	(Jq^a)_{a = 0,\ldots,2n+1}\big|_{PI} = (0,u^1,\ldots,u^n;v_0,0,\ldots,0) 
\;,
	\label{eq:JqPI}
\end{equation}
so that the PI condition can be expressed in the dual variables as
\begin{equation}
(q_a)_{a=0,\ldots, 2n+1}\big|_{PI} = - \frac{1}{H} (v_0, 0, \ldots, 0;0, -u^A) \;.
\label{eq:qPIdual} 
\end{equation}
Since the PI conditions set half of the entries in certain symplectic
vectors to zero, symplectic covariance reduces to the subgroup which
preserves this condition. We will see in the following that the equations
of motion reduce consistently to a subset of fields, provided that 
we extend the 
purely imaginary condition (\ref{eq:qPI}) to the fields
$\hat{q}^a$ by
\begin{equation}
	\left(\partial_\mu{\hat{q}}^a \right)_{a = 0,\ldots,2n+1}\Big|_{PI} = \tfrac12 \left(\partial_\mu{\zeta}^0,0,\ldots,0;0,\partial_\mu{\tilde{\zeta}}_1,\ldots,\partial_\mu{\tilde{{\zeta}}}_n\right) \;.
	\label{eq:hatqPI}
\end{equation} 
Combining expressions (\ref{eq:qPI}) and (\ref{eq:hatqPI}) we find that $q^a \Omega_{ab} \partial_\mu{q}^b = q^a \Omega_{ab} \partial_\mu{\hat{q}}^b = 0$.

We will later impose spherical symmetry on the four-dimensional
solutions, which implies that it is static. In terms of 
three-dimensional quantities staticity is equivalent to imposing
the relation (\ref{KKvectorStatic}) given below.
For static PI configurations the 
equations of motion derived from the three-dimensional Lagrangian
(\ref{L3qupstairs}), reduce to
\begin{eqnarray}
\nabla^\mu \left[\tilde{H}_{ab}\partial_\mu q^b \right] -  \partial_a \tilde{H}_{bc} \left(\partial_\mu q^b \partial^\mu q^c - \partial_\mu \hat{q}^b \partial^\mu \hat{q}^c \right) &=& 0 \nonumber\;,  \\
2 \nabla^\mu \left[\tilde{H}_{ab}\partial_\mu \hat{q}^b \right] &=& 0 \;,
\nonumber \\
\frac{1}{2} R_{\mu\nu} - \tilde{H}_{ab} \left(\partial_\mu q^a \partial_\nu q^b - \partial_\mu \hat{q}^a \partial_\nu \hat{q}^b \right)  &=& 0 \;, \label{eomPI}
\end{eqnarray}
with the Kaluza-Klein vector determined by (\ref{KKvectorStatic}).  
These equations of motion follow from the three-dimensional effective
Lagrangian 
\begin{equation}
\label{3dLPI}
e_3^{-1} {\cal L} = \frac{1}{2} R_3 - \tilde{H}_{ab} (\partial_\mu{q}^a 
\partial^\mu {q}^b
-\partial_\mu {\hat{q}}^a \partial^\mu {\hat{q}}^b ) \;,
\end{equation}
which is obtained by imposing (\ref{eq:qPI}), (\ref{eq:hatqPI})
and (\ref{KKvectorStatic})
on (\ref{L3qupstairs}).
This shows that the PI conditions represent a consistent truncation.

\subsection{Hessian metrics for PI configurations}
\label{sec:HessMetrics}

We now investigate the implications of the PI conditions 
for the Hessian metric $\tilde{H}_{ab}$. 
It is convenient to subdivide the range of the index 
$a =0, \ldots, 2n+1$ into the ranges
$\alpha, \beta, \ldots  = 0, n+2, \ldots 2n+1$ 
and $\rho, \sigma, \ldots = 1, \ldots, n+1$.  
The PI conditions 
restrict the scalar fields to the PI submanifold defined by
\[
(q^\rho) = (x^A, y_0) = 0 \;.
\]
The remaining fields
\[
(q^\alpha) = (x^0, y_A)
\]
provide coordinates for the PI submanifold. We will now show that
(\ref{eomPI}) and (\ref{3dLPI}) correspond to a sigma model
which only involves the fields $(q^\alpha, \hat{q}^\alpha)$ with 
couplings determined by the Hessian
metric $\tilde{H}_{\alpha \beta}$. 
The two non-trivial statements we have to prove are:
(i) the only surviving terms 
in the equations of motion involving
the first derivatives $\partial_a \tilde{H}_{bc}$ of the Hessian
metric are of the form $\partial_\alpha \tilde{H}_{\beta \gamma}$;
(ii) the submatrix $\tilde{H}_{\alpha \beta}$ is a Hessian
metric. The rest of this section is devoted to proving these
two statements.

From (\ref{eq:qPIdual}) we know that
the PI conditions can equivalently be written in terms of dual coordinates
\[
(q_\rho) = (\tilde{H}_\rho) = - \frac{1}{H} ( v_A, - u^0) = 0 \;,
\]
and the fields
\[
(q_\alpha) = - \frac{1}{H} ( v_0, - u^A) 
\]
provide coordinates on the PI submanifold.
The splitting of coordinates and dual coordinates into those 
tangent to the PI submanifold and those transverse to it is consistent
with our rules for raising and lowering indices, because the mixed
components of the metric vanish on the PI submanifold,
\[
\tilde{H}_{\alpha \rho} \big|_{PI}= 0  \;.
\]
More generally, since the PI condition $q^\rho=0$ implies that $\tilde{H}_\rho=
0$, it follows that any derivative of $\tilde{H}$
which contains {\em precisely
one transverse derivative}, vanishes on the PI submanifold 
$\tilde{H}_{\rho \alpha \beta \cdots}\big|_{PI}=0$ \cite{Mohaupt:2011aa}.
Next we note that the components $\tilde{H}_{\rho \sigma}$ only appear in the 
equations of motion contracted
with $\partial_\mu q^\rho$ or $\partial_\mu \hat{q}^\rho$, which vanish
if we impose the PI conditions. Moreover, since $\partial_\rho 
\tilde{H}_{\alpha \beta}\big|_{PI} =  \tilde{H}_{\rho \alpha \beta}\big|_{PI}
=0$, the only
surviving terms in the equations of motion 
involving derivatives of $\tilde{H}_{ab}$ are of the form 
$\partial_\gamma \tilde{H}_{\alpha \beta}$. Together with the vanishing
of the mixed components of the Hessian metric, 
this implies that the only remaining terms
in (\ref{eomPI}) are those where $a=\alpha=0, n+2, \ldots, 2n+1$.
We further note that $\tilde{H}_{\alpha \beta}$ is a Hessian metric
on the PI submanifold, with Hesse potential $\tilde{H}\big|_{PI}$:
\[
\tilde{H}_{\alpha \beta} \big|_{PI} = 
\left( \frac{\partial^2 \tilde{H}}{\partial q^\alpha \partial q^\beta}\right)
\big|_{PI} = 
\left( \frac{\partial^2 \tilde{H}\big|_{PI}}{\partial q^\alpha \partial q^\beta}
\right) \;,
\]
since this only involves derivatives tangential to the PI submanifold.
In the following we will use frequently that whenever a tensor component
has one index outside the range $a,b, \ldots = \alpha, \beta, \ldots = 
0, n+2, \ldots 2n+1$, 
it is either zero or decouples from the equations of motion. 
Thus we have shown that the PI conditions amount to a consistent
truncation of the three-dimensional Lagrangian to a sigma model
for the fields $(q^\alpha, \hat{q}^\alpha)$, with scalar metric 
determined by the Hessian metric $\tilde{H}_{\alpha \beta}$.\footnote{The
sigma model metric for $(q^\alpha, \hat{q}^\alpha)$
is in fact the standard para-K\"ahler metric
on the tangent bundle of the Hessian manifold parametrised by the
$q^\alpha$ \cite{2008arXiv0811.1658A,Mohaupt:2009iq}.}

\subsection{Hesse potentials for PI configurations}

It is not possible generically to compute the explicit expression for the 
Hesse potential corresponding to a prepotential of the form (\ref{eq:Ff}).
This would require solving the relation 
$\mbox{Re}\left(F_I(x^I,u^I)\right) = y_I$  to obtain $u^I=\mbox{Im}Y^I$ as
a function of $(x^I, y_I)$, which cannot be done in closed form
for a generic prepotential $F$.  
However, in this section we will show that for any 
prepotential of the form (\ref{eq:Ff}) we can find the Hesse potential explicitly as a function of the
dual variables $q_a$ after restricting to PI field configurations.
We will use the following notation for 
non-vanishing dual variables: $(q_\alpha) = (q_0, q_{\alpha'})$,
where $\alpha' = n+2, \ldots, 2n+1$. In terms of these variables,
the `PI Hesse potential' corresponding to (\ref{eq:Ff}) is
\begin{equation}
\label{Hu1}
H(q_\alpha)\big|_{PI} = - \frac{1}{2 \power +2} 
\left[ \frac{1}{\power^\power} (-q_0)^\power f(q_{\alpha'}) 
\right]^{-\frac{1}{\power +1}} \;.
\end{equation}
In the reminder of this section we will derive this formula
together with other relations that we will use later to 
solve the three-dimensional equations of motion and lift 
the solution to four dimensions.

Starting from (\ref{eq:Ff})
we compute
\begin{equation}
\label{F1st}
F_0 = -\power i^{\power -1} \frac{f(Y^1, \dots, Y^n)}{(Y^0)^{\power +1}} \;,
\;\;\;
F_A = i^{\power -1} \frac{f_A(Y^1, \dots, Y^n)}{(Y^0)^\power} \;.
\end{equation}
Next we impose the PI condition: 
\[
F(x,u)\big|_{PI} = i^{\power-1} \frac{f(iu^1, \dots, iu^n)}{(x^0)^\power} \;,
\]
\[
F_0 \big|_{PI} = iv_0 = -\power i^{2\power+1} 
\frac{f(u^1, \dots, u^n)}{(x^0)^{\power+1}} \;,\;\;\;
F_A \big|_{PI} = y_A = 
i^{2 \power} \frac{f_A(u^1, \dots, u^n)}{{x^0}^\power} \;.
\]
Here we used that $f$ is homogeneous of degree $\power+2$ and
$f_A$ homogeneous of degree $\power +1$. Note that since $f$ is 
by assumption real when evaluated on real fields, 
$f(u^1,\ldots,u^n)$ and $f_A(u^1,\ldots,u^n)$ are real homogeneous functions. 
In the following it is understood that $Y^I$, $F_I$ are subject to 
the PI condition, and we usually drop the label `PI'. The relation for $F_0$
can be used to solve for $x^0$ as a function of the dual coordinates:
\begin{equation}
\label{root}
(x^0)^{\power +1} = (-)^{\power +1} \power \frac{f(u)}{v_0} \;,
\end{equation}
where $f(u):= f(u^1, \dots, u^n)$. To obtain $x^0$, we need to
take the $(\power +1)$-st root of the above equation. Since $x^0$ must
be real, we need to distinguish two cases:
between $\power +1$ even and
$\power +1$ odd. 
\begin{itemize}
\item
If $\power +1$ is even, then $(x^0)^{\power +1}$ is positive so that
we must have $f(u) v_0 >0$. In this case the equation 
(\ref{root}) has two real roots, corresponding to $x^0>0$ and
$x^0<0$. 
\item
If $\power +1$ is odd, $(x^0)^{\power +1}$ can be positive or negative,
and we obtain no condition on $f(u) v_0$ from the reality of $x^0$. 
Moreover the
equation (\ref{root}) has a unique real root.
\end{itemize}
Thus the real solutions of (\ref{root}) are
\[
x^0 = \phi_x 
\left( \frac{\power f(u)}{v_0} \right)^{\frac{1}{\power+1}} \;,
\]
where
\[
\phi_x = \left\{ \begin{array}{ll}
\mbox{sgn}(x^0) \;, & \mbox{if} \;\;\;\power +1 \;\;\;\mbox{even} \;, \\
- 1 \;, & \mbox{if} \;\;\;\power +1 \;\;\;\mbox{odd} \;,
\end{array} \right.
\]
and where $\mbox{sgn}(x^0) =\pm 1$ is the sign of $x^0$.
We now evaluate 
\[
e^{-{\cal K}(Y)} = -i (Y^I \bar{F}_I -
F_I \bar{Y}^I) = -2 H
\] 
subject to the PI condition in order
to obtain $H(u,v)\big|_{PI}$:
\begin{eqnarray}
e^{-{\cal K}(Y)} &=& - 2 x^0 v_0 + 2 u^A y_A \nonumber \\
&=& -2 \phi_x \left[
\left( \power \frac{f(u)}{v_0} \right)^{\frac{1}{\power +1}} v_0
+ (\power+2) f(u) \left( 
\frac{v_0^\power}{\power^\power f(u)^\power} \right)^{\frac{1}{\power+1}}
\right] \;,
\end{eqnarray}
where we used that $\phi_x^\power = (-1)^{\power +1} \phi_x$ and 
$\phi_x = \phi_x^{-1}$ 
and $u^Af_A
= (\power +2) f(u)$. 
Next we move the linear factors $v_0$ and $f(u)$ inside the roots. 
If $\power +1$ is even we need to split off a factor $-1$ if $v_0$ 
is negative:
\[
v_0 = \mbox{sgn}(v_0) \left( v_0^{\power +1} \right)^{\frac{1}{\power+1}} \;,
\]
whereas for odd $\power +1$ there is no such factor. Let us therefore
define
\[
\phi_v = \left\{ \begin{array}{ll}
\mbox{sgn}(v_0) \;, & \mbox{if} \;\;\;\power +1 \;\;\;\mbox{even} \;, \\
1 \;, & \mbox{if} \;\;\;\power +1 \;\;\;\mbox{odd} \;.
\end{array} \right.
\]
We also need an analogous sign factor $\phi_f$ for $f$, but it 
turns out that $\phi_f = \phi_v$. This is clear because for
$\power+1$ odd we know that $\phi_f=1$, whereas
for $\power+1$ even we know that $f$ and
$v_0$ have the same sign.
We can thus combine terms to obtain
\[
e^{-{\cal K}(Y)} = - \phi_x \phi_v (4\power+4)  
\left( \frac{ f(u) v_0^\power}{\power^\power} \right)^{\frac{1}{\power +1}} \;.
\]
Since $e^{-{\cal K}(Y)}$ must be positive 
we obtain constraints on the signs of $f$, $v_0$ and
$x^0$. If $\power+1$ is even, the root is only real when $fv_0^\power >0$,
which is not a new condition as it is already implied by the reality
of $x^0$. Positivity of $e^{-{\cal K}(Y)}$ requires 
$\phi_x \phi_v <0$, which implies that 
$x^0$ and $v_0$ (and hence $f$) have opposite sign. If $\power+1$
is odd, then $-\phi_x \phi_v=1$ holds automatically, and we obtain $f>0$ as
the only condition.

The conditions on $f$, $v_0$ and $x^0$ can be summarised as follows

\begin{itemize}
\item
If $\power +1$ is even, then either $f(u)>0, v_0 >0$ or
$f(u) < 0, v_0<0$. Moreover the sign of $x^0$ must be 
opposite to that of $v_0$, which enters into the solution
through $\phi_x = \mbox{sgn}(x^0)$. 
\item
If $\power +1$ is odd, then $f(u) >0$, and $\phi_x = -1$.
\end{itemize}
Equivalently:
$f(u)$ and $v_0$ must satisfy
\begin{equation}\label{eq:Hreal}
v_0^\power f(u) >0 \;,
\end{equation}
and if $\power +1$ is even the signs of $v_0$ and $x^0$ must
be opposite.

For later use we note that the Hesse potential, restricted to configurations
satisfying the PI condition, is
\begin{equation}
H(u,v)\big|_{PI} = -(2\power+2)\left[\frac{1}{\power^\power} v_0^\power f \left( u^1,\ldots, u^n \right) \right]^{\frac{1}{\power+1}}  \;.
\label{eq:HPI}
\end{equation}
and the non-zero dual scalars are given by
\[
q_0 = - \frac{v_0}{H(u,v)} \;,\;\;
q_{\alpha'} = q_{A + (n+1)} = \frac{u^A}{H(u,v)} \;.
\]
Using that $f(u)$ is homogeneous of degree $\power +2$, we
can  rewrite (\ref{eq:HPI}) in terms of $q_a$:
\begin{equation}
\label{Hu}
H(u(q_\alpha), v(q_\alpha))\big|_{PI} = - \frac{1}{2 \power +2} 
\left[ \frac{1}{\power^\power} (-q_0)^\power f(q_{\alpha'}) 
\right]^{-\frac{1}{\power +1}} \;.
\end{equation}
For notational convenience we
will set $H(q_\alpha) := H(u(q_\alpha), v(q_{\alpha}))\big|_{PI}$ 
in the following.\footnote{In the notation of section \ref{sec:dualcoords}
the correct notation would be $H''(q_a)$. Note that $H(q_a)$ is not
an invariant function under the diffeomorphism 
$(u^I,v_I) \mapsto q_a$.}

Let us explain how to check that this expression is real and negative,
as required. Similar arguments can be used as quick checks for the
correctness of the various explicit solutions we give later. All we 
need to do is to re-write the conditions
(\ref{eq:Hreal}) in terms of the dual variables $q_a$. First note
that since $H<0$ and $q_0 = - H^{-1} v_0$, it follows that 
$q_0$ and $v_0$ have the same sign. Next, $q_{A+(n+1)} = H^{-1} u^A$,
so that $q_{A+(n+1)}$ and $u^A$ have opposite signs. Since
$f(u) = H^{-(\power +2)} f(q_{\alpha'})$, where $\alpha' = A+(n+1)$,
it follows that $f(q_{\alpha'})$ has the same (opposite) sign
to $f(u)$ if $\power +1$ is odd (even). Thus the conditions
for consistent real solutions are:

\vspace{.5em}
\noindent
$q_0$ and $f(q_{\alpha'})$ must satisfy
\[
(-q_0)^\power f(q_{\alpha'}) > 0 \;.
\]
If $\power +1$ is even, then the sign of $x^0$ must be 
opposite to that of $q_0$, which enters into solutions 
through $\phi_x = \mbox{sgn}(x^0)$. If $\power +1$ is odd, then
$\phi_x = -1$. 
\vspace{.5em}

From this criterion it is manifest that $H(q_a)$ as given in (\ref{Hu})
is real and negative.

Note that even for purely imaginary field configurations it is 
still not possible to 
find an explicit expression for the Hesse potential in terms of $(x^I,y_I)$ 
without imposing further conditions. One class where this is possible
are the diagonal models, which will be discussed in section (\ref{genmodelsec}).

\subsection{Spherical symmetry}

Besides the PI condition we impose
that all four-dimensional fields (metric, scalars and gauge fields) are spherically symmetric. Spherically symmetry spacetime metrics are reviewed in appendix
\ref{app:A}.
According to (\ref{eq:MetricTau}) the three-dimensional part of any four-dimensional stationary and spherically symmetric spacetime metric can be written in the form
\[
	g^{(3)} = e^{4{\cal A}(\tau)} d\tau^2 + e^{2{\cal A}(\tau)} d\Omega_{(2)}^2 \;.
\]
In terms of the radial coordinate $\tau$ the three-dimensional Laplacian takes the simple form $\Delta = \frac{d^2}{d\tau^2}+ \cdots$, where the omitted terms are independent of $\tau$. The solution for the scalar fields corresponds to
a geodesic curve in the scalar manifold $\bar{N}$ of
the three-dimensional theory. The advantage of the radial coordinate
$\tau$ compared to other (not affinely related) choices of a radial
coordinate is that $\tau$ provides an affine parametrisation of this
geodesic.

It turns out that the three-dimensional Einstein equations completely fix the function ${\cal A}(\tau)$. Discarding solutions that are periodic in $\tau$ one finds that $e^{-{\cal A}(\tau)} = \frac{\sinh c\tau}{c},$ for some constant $c$. 
The three-dimensional metric then takes the form \cite{Breitenlohner:1987dg}
\begin{equation}\label{eq:3dmetric}
g^{(3)} = \frac{c^4}{\sinh^4 c\tau} d\tau^2 + \frac{c^2}{\sinh^2 c\tau} d\Omega_{(2)}^2 \;,
\end{equation}
which is precisely the three-dimensional part of the Reissner-Nordstr\"om 
metric. 

For the interpretation as a dimensionally reduced black hole, 
it is convenient to 
replace the `affine' radial coordinate $\tau$, 
by a different radial coordinate $\rho$, defined by the relation 
\begin{equation}
\label{W}
W(\rho): = 1-\frac{2c}{\rho}=e^{-2c \tau} \;,
\end{equation}
in which case
\[
g^{(3)} = \frac{d\rho^2}{W} + \rho^2 d \Omega^2_{(2)} \;.
\]
The parameter $c\geq 0$ is the non-extremality parameter, with
$c=0$ being the extremal limit. The outer horizon is at $\rho=2c$,
which corresponds to $\tau \rightarrow \infty$. Using the radial
coordinate $\rho$, the solution can be continued analytically from 
the outer horizon to the inner horizon located at $\rho=0$.

Combining the fact that spacetime is both spherically symmetric and stationary is enough to ensure that is is static, the proof of which is reviewed in 
appendix \ref{app:A}.
Therefore one may choose coordinates in which the KK-vector vanishes
in (\ref{eq:KKdecomp}). In terms of three-dimensional fields this means that  
\begin{equation}
\label{KKvectorStatic}
	\frac{1}{2H}\left( \partial_\mu \tilde{\phi} +\tfrac{1}{2}(\zeta^I \partial_\mu \tilde{\zeta}_I - \tilde{\zeta}_I \partial_\mu \zeta^I)\right) = \frac{1}{2H} \left( \partial_\mu \tilde{\phi} + 2 \hat{q}^c \Omega_{cd} \partial_\mu \hat{q}^d \right) = 0 \;.
\end{equation}
This term (squared) appears in isolation in the Lagrangian (\ref{L3qupstairs})
and therefore decouples from all other terms in the equations of motion. 
An effective Lagrangian for general static configurations
is given by taking the first two lines of (\ref{eq:3dLag}). If we impose in
addition the PI conditions, then this reduces to the first line,
with half the scalar fields being constant, as discussed previously in 
this section. If we impose spherical symmetry together with the PI 
conditions, then staticity is implied, and all fields can be taken
to only depend on the affine radial coordinate $\tau$ using
the parametrisation (\ref{eq:3dmetric}).
Then the equations (\ref{eomPI}) reduce to 
\begin{eqnarray}
\frac{d}{d\tau} \left( \tilde{H}_{ab} \dot{q}^b
\right) 
-  \partial_a \tilde{H}_{bc} \left(\dot{q}^b \dot{q}^c - \dot{\hat{q}}^b 
\dot{\hat{q}}^c \right) &=& 0 \;, \nonumber \\
\frac{d}{d\tau} \left( \tilde{H}_{ab} \dot{\hat{q}}^b\right) &=& 0 \;, \nonumber \\
\tilde{H}_{ab} 
\left(\dot{q}^b \dot{q}^c - \dot{\hat{q}}^b 
\dot{\hat{q}}^c \right)
  &=& c^2 \;, \label{eomPIspherical}
\end{eqnarray}
where a dot denotes the differentiation with respect to $\tau$. 
The first two equations are the scalar equations of motion, which 
are equivalent to the geodesic equation for the curve 
$(q^a(\tau), \hat{q}^a(\tau))$ 
on the scalar manifold $\bar{N}$.
The scalar equations of motion 
follow from the one-dimensional effective Lagrangian
\begin{equation}
		{\cal L}_1 = -\tilde{H}_{ab} \left(\dot{q}^a \dot{q}^b -  \dot{\hat{q}}^a \dot{\hat{q}}^b \right) \;.
\label{eq:3deffLag} 
\end{equation}
The third equation of (\ref{eomPIspherical}), 
which is the non-trivial component of the higher-dimensional
Einstein equations, is the Hamiltonian constraint which needs
to be imposed on top of the one-dimensional Euler-Lagrange 
equations.

In our applications it will be convenient to use the inverse metric 
$\tilde{H}^{ab}$ 
and the dual coordinates $q_a$. To perform the rewriting we
use the relations
(\ref{qupqdown}) and (\ref{tildeHderivativereln}), 
and the relation 
$\partial_d \tilde{H}^{ab} = - \tilde{H}^{ac} \tilde{H}^{be}
\partial_d \tilde{H}_{ce}$ between the first derivatives of a metric
and those of its inverse. Note that 
indices on the vector fields $\dot{q}^a$, $\dot{\hat{q}}^a$ and derivatives
$\partial_a=\frac{\partial}{\partial q^a}$ are
raised and lowered with $\tilde{H}_{ab}$,  in particular that 
$\tilde{H}^{ab} \frac{\partial}{\partial q^b} = \frac{\partial}{\partial q_a}$.
In terms of the dual variables, 
the 
scalar equations of motion are
\begin{equation}\label{qeom}
\ddot{q}_a + \frac{1}{2} \tilde{H}_{ad} \partial^d \tilde{H}^{bc} \left( \dot{q}_a \dot{q}_b - \dot{\hat{q}}_a \dot{\hat{q}}_b \right) = 0\;,
\end{equation}
and
\begin{equation}\label{qhateom}
\ddot{\hat{q}}_a =0 \;,
\end{equation}
and the Hamiltonian constraint takes the form
\begin{equation}\label{hamconstraint}
\tilde{H}^{bc} \left(  \dot{q}_a \dot{q}_b - \dot{\hat{q}}_a \dot{\hat{q}}_b 
\right) =c^2 \;,
\end{equation}
where $\dot{\hat{q}}_a := \tilde{H}_{ab} \dot{\hat{q}}^b$ is the
co-vector field obtained by lowering the index of the vector
field $\dot{\hat{q}}^a$. 

We remark that we do not require the existence of `dual coordinates'
$\hat{q}_a$ as functions on the scalar manifolds.
In particular it is not possible to define
dual coordinates as $\tilde{H}_{ab} \hat{q}^b$ (unless
$\tilde{H}_{ab}$ is constant), because this would not 
be consistent with $\dot{\hat{q}}_a = \tilde{H}_{ab} \dot{\hat{q}}^b.$
However $\hat{q}^a$ are well defined functions on the scalar manifold, 
and $\dot{\hat{q}}^a$ and $\dot{\hat{q}}_a$ are well defined
(co-)vector fields.

The remainder of this paper is dedicated to solving the equations of motion written in the dual coordinates (\ref{qeom}) - (\ref{hamconstraint}). It is worth reiterating that we have only imposed that solutions are stationary, spherically symmetric and purely imaginary. Recall that the latter condition means that 
\[
	q_\rho = \dot{\hat{q}}_\rho = 0 \;, \qquad \rho = 1, \ldots, n + 1 \;,
\]
and, as discussed in \ref{sec:HessMetrics}, this implies that the equations of motion (\ref{qeom}) - (\ref{hamconstraint}) only involve the fields 
\[
	(q_\alpha, \dot{\hat{q}}_\alpha) \;, \qquad \alpha = 0, n+2, \ldots 2n + 1 \;.
\]
With this in mind, the $\dot{\hat{q}}_a$ equation of motion can be immediately integrated to give 
\begin{equation}\label{eq:hatqdot}
\dot{\hat{q}}_a = K_a = \left( - \mathcal{Q}_0,0,\ldots,0;0,\mathcal{P}^1,\ldots,\mathcal{P}^n \right) \;,
\end{equation}
where the integration constants $\mathcal{Q}_0, \mathcal{P}^A$ are proportional to the electric and magnetic charges of the black hole solution.\footnote{The 
minus sign in front of ${\cal Q}_0$ is included in view of the relation
$q_a = \frac{1}{H} (- v_I, u^I)$. Our sign conventions are such 
that for BPS solutions the attractor equations take the same form
as in \cite{LopesCardoso:2000qm}.} The $n + 1$ charges $\mathcal{Q}_0, \mathcal{P}^A$ are the maximum number allowed for purely imaginary configurations, and they may be freely chosen for all solutions considered in this paper, regardless of the model in question.

\section{Three-dimensional instanton solutions}
\label{sec:3dsols}

We will now construct explicit solutions to the equations
(\ref{qeom}) - (\ref{hamconstraint}), which we refer to as instanton
solutions.

\subsection{Instanton solutions for diagonal models}
\label{genmodelsec}

We start by discussing a class of models where we will
be able to find the general purely imaginary solution in closed
form. The prepotential is restricted to have the form
\begin{equation}\label{eq:Fsclass}
F = i^{\power-1} \frac{(Y^1\ldots Y^n)^{\frac{\power + 2}{n}}}{(Y^0)^\power} \;.
\end{equation}
For reasons that will become clear 
we refer to this class as  
\emph{diagonal} models. They form a two-parameter family parametrised
by $\power, n=1,2,3,\ldots$. The particular choice $\power=1, n=3$ 
corresponds to the well-known $STU$ model. We will see that the family
of diagonal models shares certain features of the $STU$-model, in particular
they allow for explicit solutions, although such models do in 
general not correspond to homogeneous spaces. 

According to (\ref{Hu1}) after imposing the PI conditions we can write the Hesse potential for this class of models as 
\begin{equation}
\label{Hqsuba}
H(q_a) 
=-\frac{1}{2\power+2} \left[ \left( \frac{-q_0}{\power} \right)^\power  \left(q_{n+2} \dots q_{2n+1} \right)^{\frac{\power+2}{n}} \right]^{-\frac{1}{\power+1}} \;.
\end{equation} 
This is manifestly real and negative for 
$(-q_0)^\power \left(q_{n+2} \dots q_{2n+1} \right)^{\frac{\power+2}{n}} >0$,
which is (\ref{eq:Hreal}) expressed in terms of $q_a$.
Both the $q_a$ equation of motion (\ref{qeom}) and the Hamiltonian constraint 
(\ref{hamconstraint}) require us to compute the matrix $\tilde{H}^{ab}$ given by
\begin{equation}\label{tildeHabformula}
\tilde{H}^{ab} = \frac{1}{2H} \frac{\partial^2H}{\partial q_a \partial q_b} 
- \frac{1}{2H^2}
\frac{\partial H}{\partial q_a} \frac{\partial H}{\partial q_b}\;, \;\;\;
H=H(q_a) \;,
\end{equation}
which follows from (\ref{tildeHabformula1}) by setting $C=-\frac{1}{2}$.

Before entering into explicit calculations, we can already observe
that for prepotentials of the from (\ref{eq:Fsclass}) the Hessian metric
for PI field configurations exhibits further simplifications 
compared to the general class (\ref{eq:Ff}). By taking the 
logarithm of (\ref{Hqsuba}) we obtain
\begin{equation}\label{eq:tildeHlogsum}
\tilde{H} \sim \log{\left(q_0\right)}^{-\frac{\power}{\power+1}} + \log{\left(q_{n+2}\right)}^{-\frac{\power+2}{(\power+1)n}} + \dots + \log{\left(q_{2n+1}\right)}^{-\frac{\power+2}{(\power+1)n}}  \;,
\end{equation}
from which it is easy to see that applying (\ref{tildeHabformula}) leads to 
a matrix with the following block structure
\begin{equation}\label{blockdecomp}
\tilde{H}^{ab} =  \frac{\partial^2 ( -\tilde{H} )}{\partial q_a \partial q_b} 
= \left( \begin{array}{c|ccc|ccc}
 \tilde{H}^{00} & & 0 & & & 0 & 
 \\ \hline 
     &  *  & \dots  &  *   &   &   &  
 \\ 0  & \vdots & \ddots & \vdots & & 0 & 
 \\ &  *  & \dots  & *   &   &  & 
 \\ \hline  \vphantom{\overset{=}{H}}  &        &        &        & \tilde{H}^{n+2,n+2} &  &  
 \\ 0 & & 0 &  &  &  \ddots &  
 \\  &  &  &  &  &  & \tilde{H}^{2n+1,2n+1}     \end{array} \right) \;.
\end{equation}
The central $(n+1) \times (n+1)$ block $\tilde{H}^{\rho \sigma}$ 
contains unknown and potentially nonzero entries that we represent with a 
`$*$.' 
However, we have shown that for PI field configurations this block decouples
from the equations of motion. We also observe the vanishing of mixed entries
of the form $\tilde{H}^{\alpha \rho}$, as derived previously in generality. 
The additional simplification, which is obvious from the fact that 
$\tilde{H}$ given in (\ref{eq:tildeHlogsum}) is a sum of terms each 
depending on precisely one coordinate,
is that the submatrix $\tilde{H}^{\alpha \beta}$ is diagonal. This phenomenon,
which motivates the terminology `diagonal models'
was already observed in \cite{Mohaupt:2009iq}
for five-dimensional 
extremal black holes, and \cite{Mohaupt:2011aa} for four-dimensional
extremal black holes and in \cite{Dempster:2013mva} for extremal and
non-extremal black strings.

Using (\ref{tildeHabformula}), we find the nonzero entries relevant for the equations of motion, and their derivatives, to be
\begin{alignat}{4}
\tilde{H}^{00} &=\frac{\power}{2\power+2} q_0^{-2} \;, &\partial^0 \tilde{H}^{00} &= -2 \frac{\power}{2\power+2} q_0^{-3} \label{eq:tildeHab1} \;,
\\ \tilde{H}^{n+2,n+2} &= \frac{\power+2}{(2\power+2)n} q_{n+2}^{-2} \;, &\partial^{n+2} \tilde{H}^{n+2,n+2} &= -2 \frac{\power+2}{(2\power+2)n} q_{n+2}^{-3}  \notag \;,
\\
\vdots \notag
\\
\tilde{H}^{2n+1,2n+1} &= \frac{\power+2}{(2\power+2)n} q_{2n+1}^{-2} \;, \quad &\partial^{2n+1} \tilde{H}^{2n+1,2n+1} &= - 2 \frac{\power+2}{(2\power+2)n} q_{2n+1}^{-3} \label{eq:tildeHab2} \;.
\end{alignat}
Note that each diagonal matrix element only depends on the
corresponding scalar field, thus leading to a complete decoupling of the
scalar equations of motion. 
Because of the diagonal structure of $\tilde{H}^{ab}$, the inverse elements, $\tilde{H}_{ab}$, of the above entries are easy to obtain e.g.
\begin{equation}
\tilde{H}_{00} = \left( \tilde{H}^{00} \right)^{-1} = \frac{1}{\tilde{H}^{00}} = \frac{2\power+2}{\power} q_0^2 \;,
\end{equation}
and similarly for the other components. We can use this to compute the quantities $\frac{1}{2} \partial_a \tilde{H}^{bc} = \frac{1}{2} \tilde{H}_{ad} \partial^d \tilde{H}^{bc}$ that appear in the equations of motion
\begin{align}\label{eq:qeomfactor}
\frac{1}{2} \partial_0 \tilde{H}^{00} &= - q_0^{-1}
\\
\frac{1}{2} \partial_{n+1+A} \tilde{H}^{n+1+A,n+1+A} &= - q_{n+1+A}^{-1} \;,
\end{align}
where $A=1, \dots, n$.

Replacing $\dot{\hat{q}}_a$ by $K_a$ according to (\ref{eq:hatqdot}), the $q_a$ equation of motion (\ref{qeom}) becomes
\begin{equation}
\ddot{q}_a + \frac{1}{2} \tilde{H}_{ad} \partial^d \tilde{H}^{bc} \left( \dot{q}_b \dot{q}_c - K_b K_c \right) =0 \;.
\end{equation}
Substituting from (\ref{eq:qeomfactor}) and (\ref{eq:hatqdot}), we see that the individual equations look like
\begin{alignat}{4}
a&=0 \;,  &\ddot{q}_0 - q_0^{-1} \left( \dot{q}_0^2 - \mathcal{Q}_0^2 \right) &= 0 \;, \notag
\\ a&=n+2 \;, &\ddot{q}_{n+2} - q_{n+2}^{-1} \left( \dot{q}_{n+2}^2 - (\mathcal{P}^1)^2 \right) &= 0 \;, \notag
\\ & \hspace{1cm} \vdots   & & \notag
\\ a&=2n+1 \;, \qquad &\ddot{q}_{2n+1} - q_{2n+1}^{-1} \left( \dot{q}_{2n+1} - (\mathcal{P}^n)^2 \right) &= 0 \;. \label{eq:2ndOrderEOM}
\end{alignat}
These equations are solved by 
\begin{align}\label{eq:qdowninstsoln}
q_0 &= \pm  \frac{-\mathcal{Q}_0}{B_0} \sinh{ \left( B_0 \tau + B_0 \frac{h_0}{\mathcal{Q}_0} \right)} \;, \notag
\\ q_{n+2} &= \pm \frac{\mathcal{P}^1}{B^1} \sinh{ \left( B^1 \tau + B^1 \frac{h^1}{\mathcal{P}^1} \right)} \;, \notag
\\ &\vdots \notag
\\ q_{2n+1} &= \pm \frac{\mathcal{P}^n}{B^n} \sinh{\left( B^n \tau + B^n \frac{h^n}{\mathcal{P}^n} \right)} \;,
\end{align}
where $B_0,B^A,h_0,h^A$ are integration constants. Since making the replacement $B_0, B^A \mapsto -B_0, -B^A$ leaves the solution invariant we may assume without loss of generality that the integration constants $B_0, B^A$ are non-negative.
The choice of sign distributions in (\ref{eq:qdowninstsoln}) has an interesting effect when lifting to four-dimensional black holes: when taking the extremal limit one obtains BPS black holes for the case where all signs are equal, whereas for all other sign distributions one obtains non-BPS black holes. 
We will not address this further in the present paper, but
refer the reader to \cite{Mohaupt:2011aa} for more information on this 
topic.
For convenience we will choose the positive sign in the above expressions from now on. 

Having eliminated $\dot{\hat{q}}_a$ by their equation of motion the
Hamiltonian constraint (\ref{hamconstraint}) becomes a condition on the 
the scalar fields $q_a$. We can use (\ref{eq:tildeHab1}) - (\ref{eq:tildeHab2}) 
and (\ref{eq:hatqdot}) to expand this as
\begin{align}
&\frac{\power}{2\power+2} q_0^{-2} \left( \dot{q}_0^2 - K_0^2 \right) 
+ \frac{\power+2}{(2\power+2)n} q_{n+2}^{-2} \left( \dot{q}_{n+2}^2 - K_{n+2}^2 \right) \notag
\\ + &\dots + \frac{\power+2}{(2\power+2)n} q_{2n+1}^{-2} \left( \dot{q}_{2n+1}^2 - K_{2n+1}^2 \right) = c^2 \;. \label{eq:hamconstnosub}
\end{align}
Substituting our solution for the scalars $q_a$ into this, we see the Hamiltonian constraint becomes
\begin{equation}\label{eq:hamconstsub}
\frac{\power}{2\power+2} \left( B_0 \right)^2 + \frac{\power}{(2\power+2)n} \left( B^1 \right)^2 + \dots + \frac{\power}{(2\power+2)n} \left( B^n \right)^2 = c^2 \;,
\end{equation}
which can be viewed either as a constraint on the integration constants $B_0,B^A$ or on the non-extremality parameter, $c$.

The instanton solution for $q_a$ and $\dot{\hat{q}}_a$ given 
in (\ref{eq:qdowninstsoln}) and (\ref{eq:hatqdot}) respectively, subject to the Hamiltonian constraint (\ref{eq:hamconstsub}), 
is general in the sense that for 
$2(n+1)$ independent scalar fields $q_\alpha, \hat{q}_\alpha$,
$\alpha=0, n+2, \ldots 2n+1$ subject to second order field equations
we have $4(n+1)$ integration constants. These may  be counted as follows: 
if we regard $c$ as a dependent quantity, then $2(n+1)$ integration 
constants are given by $B_0, B^A, h_0, h^A$ appearing in the
solution of the $q_a$ equation of motion. The charges ${\cal Q}_0, {\cal P}^A$ provide a further $n+1$ integration 
constants. The remaining $n+1$ integration constants, which 
are obtained by integrating $\dot{\hat{q}}_a = K_a$ are 
unphysical due to the axionic shift symmetries of the fields
$\hat{q}^a$, which reflect the four-dimensional gauge symmetry. 

We remark that while we have not defined dual coordinates
$\hat{q}_a$ as functions on the scalar manifold, one can of course
integrate $\dot{\hat{q}}_a = K_a$ along the curve representing
the solution, and thus obtain functions $\hat{q}_a(\tau) =
K_a \tau + R_a$ along
that curve. Alternatively, $\hat{q}^a$ are well defined functions
on the scalar manifold, an 
integration  of $\dot{\hat{q}}^a
= \tilde{H}^{ab} K_b$ will involve $n+1$ integration constants.
However these integration constants will drop out of any 
four-dimensional gauge invariant quantity, so that only $3(n+1)$ 
integration constants are relevant. We will see later that
four-dimensional black hole regularity conditions reduce this further to
$2(n+1)$ integration constants, which reflects the existence of
a unique first order rewriting of the $q^a$ equations of motion.

We further remark that using the explicit expressions
(\ref{eq:tildeHab1}) -- (\ref{eq:tildeHab2}) we can 
obtain an explicit expression for the Hesse potential
$H(x,y)\big|_{PI}$ in terms of the special real variables
$q^a=(x^I,y_I)$, restricted to PI configurations:
\[
H(q^a)\big|_{PI} 
= C \left[ (-q^0)^\power (q^{n+2} \cdots q^{2n+1})^{\frac{\power+2}{n}}
\right]^{\frac{1}{\power +1}} \;,
\]
where $C$ is a numerical constant that does not enter into 
the expression $\tilde{H}_{ab}$. Note that by expressing $q_a$ 
in terms of $q^a$ the power $-\frac{1}{\power+1}$ gets replaced
by its negative.

\subsection{The universal instanton solution \label{sec:universal}}

The opposite case to a diagonal model is a model where $\tilde{H}^{ab}$,
after imposing the PI conditions, 
does not admit a further block decomposition, so that every non-vanishing scalar field 
$q_\alpha$ couples with all others. In this case we can still find a solution
with one independent three-dimensional 
scalar field by taking the fields $q_\alpha$ to
be proportional to each other
\[
q_\alpha =\propconst_\alpha q \;,
\]
where the constants $\propconst_\alpha$ will turn out to be determined by
the charges. 
For $q$ we take the same solution as for scalars in diagonal 
models,
\begin{equation}
\label{q-equation}
q =\pm \frac{K}{B} \sinh (B\tau + \frac{Bh}{K}) \;.
\end{equation}
Thus $q$ satisfies
\[
\ddot{q} = \frac{\dot{q}^2 - K^2}{q} = B^2 q \Rightarrow
\frac{\dot{q}^2 -K^2}{q^2} = B^2 \;.
\]
Note that $\ddot{q}_\alpha = B^2 q_\alpha$. The homogeneity properties
of the Hessian metric imply \cite{Mohaupt:2010fk}
\[
q_\alpha = - \tilde{H}_{\alpha\beta} q^\beta =\frac{1}{2} q^\gamma \partial_\gamma \tilde{H}_{\alpha\beta} q^\beta
=\frac{1}{2} \partial_\alpha \tilde{H}_{\beta\gamma} q^\beta q^\gamma 
= - \frac{1}{2} \partial_\alpha \tilde{H}^{\beta\gamma} q_\beta q_\gamma
= - \frac{1}{2} \tilde{H}_{\alpha\delta} \partial^\delta \tilde{H}^{\beta\gamma} q_\beta q_\gamma \;.
\]
Using this when substituting back (\ref{q-equation}) into the 
$q_a$ equation of motion we obtain
\[
\tilde{H}_{\alpha\delta} \partial^\delta \tilde{H}^{\beta\gamma}
\left( K^2 \propconst_\beta \propconst_\gamma - K_\beta K_\gamma \right) = 0 \;.
\]
This can be solved by imposing the constraint
\[
\propconst_\alpha = \frac{K_\alpha}{K} \;,
\]
which fixes the constants of proportionality between the scalars
$q_\alpha$ in terms of the charges $K_\alpha$, up to the overall scale
$K$, which drops out of ratios:
\[
\frac{\propconst_\alpha}{\propconst_\beta} = \frac{q_\alpha}{q_\beta} = \frac{K_\alpha}{K_\beta} \;.
\]
It remains to solve the Hamiltonian constraint. Here we use that
$\tilde{H}^{\alpha\beta}$ is homogeneous of degree $-2$ in the variables $q_\alpha$:
\[
\tilde{H}^{\alpha\beta}(q_\alpha) = q^{-2} \tilde{H}^{\alpha\beta}(\propconst_\alpha) \;.
\]
Then the Hamiltonian constraint becomes
\[
\frac{\tilde{H}^{\alpha\beta}(\propconst)}{q^2} \left( \propconst_\alpha \propconst_\beta 
\dot{q}^2 - K_\alpha K_\beta \right) = 
B^2 \tilde{H}^{\alpha\beta}(\propconst) \propconst_\alpha \propconst_\beta = c^2 \;.
\]
This is an algebraic constraint which fixes $B$ in terms of
$c$ and the charges. We will see in Section \ref{genliftsec}
that the universal solution corresponds to a four-dimensional
solution with the non-extremal Reissner-Nordstr\"om metric, 
multiple charges, and constant four-dimensional scalars.

\subsection{Instanton solutions for block diagonal models}
\label{blockinstantonsec}

In this section we explain how to obtain explicit instanton solutions 
for non-diagonal models, assuming that $\tilde{H}^{\alpha \beta}$ 
decomposes into two or more blocks. We will show that in this case
we can obtain explicit solutions which still carry all the 
gauge charges
consistent with the PI conditions, but with a reduced number 
of independent scalar fields, because the solutions for scalar
fields belonging to the same block will be proportional, with
ratios determined by the ratios of the corresponding gauge 
charges.

To keep formulas simple we will only 
consider the case $\power=1$, with 
prepotentials of the form
\begin{equation}
\label{eq:genverspec}
F = \frac{f(Y^1, \ldots, Y^n)}{Y^0} \;,
\end{equation}
with $f(Y^1, \ldots, Y^n)$ homogeneous of degree 3,
and real when evaluated on real fields. 
The corresponding Hesse potential for PI configurations is
\[
H(q_0, q_{\alpha'}) = - \frac{1}{4} \left[
(-q_0) f(q_{\alpha'}) \right]^{-\frac{1}{2}} \;.
\]
Note that all `very special' prepotentials
that can be obtained by dimensional reduction are of this type. 
As observed in \cite{Mohaupt:2011aa},
when imposing the PI conditions it follows that $\tilde{H}^{0 \beta'}=0$ for
$\beta'=n+2, \ldots, 2n+1$, so that $\tilde{H}^{\alpha \beta}$ always
subdivides into at least two blocks, $\tilde{H}^{00}$ and a further
$n\times n$ block $\tilde{H}^{\alpha' \beta'}$:
\begin{equation}\label{blockdecomp2}
\tilde{H}^{ab} =  \frac{\partial^2 ( -\tilde{H} )}{\partial q_a \partial q_b} 
= \left( \begin{array}{c|ccc|ccc}
 \tilde{H}^{00} & & 0 & & & 0 & 
 \\ \hline 
     &  *  & \dots  &  *   &   &   &  
 \\ 0  & \vdots & \ddots & \vdots & & 0 & 
 \\ &  *  & \dots  & *   &   &  & 
 \\ \hline  \vphantom{\overset{=}{H}}  &        &        &        
& \tilde{H}^{n+2,n+2} & \cdots  & \tilde{H}^{n+2, 2n+1}  
 \\ 0 & & 0 &  & \vdots &  \ddots &  \vdots 
 \\  &  &  &  & \tilde{H}^{2n+1,n+2} &\cdots  & \tilde{H}^{2n+1,2n+1}     
\end{array} \right) \;.
\end{equation}

If one restricts
the form of $f(Y^1, \ldots, Y^n)$ then $\tilde{H}^{\alpha'\beta'}$
might decompose into further blocks,\footnote{One might of course need
to perform row operations to make the decomposition explicit.}
the limiting case being diagonal models.

To be precise, a block decomposition of the equations of motion 
does not only require that $\tilde{H}^{\alpha \beta}$ exhibits a block 
structure. The full set of conditions 
is obtained in the same way as when we discussed the consistent
truncation of the equations of motion by the PI condition
in Section \ref{SphericalPI}. To have
a decoupling one also needs that the matrix elements in each block 
only depend on the scalar fields corresponding to this block.
This implies in particular that the derivatives 
$\partial^{\gamma} \tilde{H}^{\alpha \beta}$ exhibit the same block
decomposition as $\tilde{H}^{\alpha \beta}$ itself. 
For terminological convenience
we will refer to these conditions as $\tilde{H}^{\alpha \beta}$ `admitting
a block decomposition.' The conditions are met for 
(\ref{eq:genverspec}), and all the further examples that we will
discuss. It is clear 
that a block decomposition always occurs if
$H$ is a product with factors depending on disjoint subsets of variables,
so that $\tilde{H}$ is a sum of terms depending on disjoint subsets 
of variables, which implies that $\tilde{H}^{\alpha \beta}$ is a product
metric.\footnote{Further examples can arise whenever the further
consistent truncation of a model induces a decoupling of the 
field equations for the remaining fields. 
We will not investigate
this systematically in the present paper.}

One important class of examples which always allows a further
block decomposition are prepotentials of the form
\begin{equation}
\label{eq:2blockprepot}
F = \frac{f_1(Y^1) f_2(Y^2, \ldots, Y^n)}{Y^0} \;.
\end{equation}
This class contains
tree-level heterotic prepotentials, which 
are always linear in the dilaton $Y^1/Y^0$, 
${\cal N}=2$ truncations of ${\cal N}=4$ theories,
and models based on reducible Jordan algebras.
The corresponding Hesse potential for PI configurations is
\[
H(q_0, q_{n + 2}, q_{n + 3}, \ldots ) = - \frac{1}{4} 
\left[ (-q_0) f_1(q_{n+2}) f_2(q_{n+3}, q_{n+4}, \ldots) \right]^{-\frac{1}{2}} \;,
\]
so that 
\begin{equation}\label{Habblockdecomp}
\tilde{H}^{ab} = \left( \begin{array}{c|ccc|cccc} 
\frac{1}{4} q_0^{-2} & 0 & \cdots & 0 & 0 & 0 & 0 & \cdots\\ \hline
		0	& * & \cdots & * & 0 & 0 & 0 & \cdots\\
		\vdots	& \vdots &   & \vdots & 0 & 0 & 0 & \cdots\\
		0	& * & \cdots & * & 0 & 0 & 0 & \cdots\\ \hline
		0 & 0 & 0 & 0 & * & 0 & 0 & \cdots\\		
		0 & 0 & 0 & 0 & 0 & * & * & \cdots \\		
		0 & 0 & 0 & 0 & 0 & * & * & \cdots	\\	
		\vdots & \vdots & \vdots & \vdots & \vdots & \vdots & \vdots & \ddots		
		\end{array} \right) \;.
	\end{equation}

Whenever a model is not diagonal its scalar fields will couple, which
will prevent us from finding the general solution by the method used
in the previous section. However, if $\tilde{H}^{\alpha \beta}$ has
a block decomposition, then only the scalars corresponding to the
same block couple to one another. One can then proceed by taking
all scalar fields belonging to the same block to be proportional. 
In this case the method described in Section \ref{sec:universal} gives
a non-trivial solution, though not the most general one since one
only has as many independent scalar fields as one has blocks. 

Let us assume that there are $M$ blocks, labeled by $m=1, \ldots, M$. 
For each block we take all the corresponding scalars to be proportional
to 
\[
q_{(m)} = \pm \frac{K^{(m)}}{B^{(m)}} \sinh \left( B^{(m)} \tau + 
\frac{B^{(m)} h^{(m)}}{K^{(m)}} \right) \;.
\]
Since the blocks decouple this solves the $q_a$ equations of motion
for the $m$-th block, with constants of proportionality fixed 
by the corresponding charges:
\[
\propconst_a^{(m)} = \frac{K_a^{(m)}}{K^{(m)}} \;,\;\;\;
\frac{\propconst_a^{(m)}}{\propconst_b^{(m)}} = 
\frac{q_a^{(m)}}{q_b^{(m)}} = 
\frac{K_a^{(m)}}{K_b^{(m)}} \;,
\]
where the indices $a,b,\ldots$ are restricted to values corresponding
to the $m$-th block. In the following we will omit the superscript 
$(m)$ on $\propconst_a$ and $K_a$ whenever it is clear to which block 
they belong. 

The Hamiltonian constraint (\ref{hamconstraint})
couples the scalars in different blocks:
\begin{equation}
\sum_{m=1}^M \sum_{a,b \in I_{(m)}} \tilde{H}^{ab}(\propconst)
\frac{ \propconst_a \propconst_b \dot{q}_{(m)}^2 - K_a K_b}{q_{(m)}^2} =
\sum_{m=1}^M (B^{(m)})^2 \psi_m  = c^2 \;,
\end{equation}
where
\begin{equation}
\label{psim}
\psi_m = \sum_{a,b \in I_{(m)}} \tilde{H}^{ab}(\propconst) \propconst_a \propconst_b\;,
\end{equation}
with $I_{(m)}$ the subset of indices corresponding to the $m$-th block.  
We remark that we will see in Section \ref{genblockliftsec}
that for regular black hole solutions
$B^{(m)}= c$ for all $m$, so that  the condition
\begin{equation}
\label{SumPsi}
\sum_{m=1}^M \psi_m =1 
\end{equation}
must be satisfied.

For prepotentials of the form (\ref{eq:Ff}), subject to the PI conditions, 
there will always be a 
single $1\times 1$ block corresponding to the field $q_0$. A decomposition
of the complementary block $q_{n+2}, \ldots q_{2n+1}$ will occur for special
choices of the function $f(Y^1, \ldots, Y^n)$, as for the example
given by (\ref{eq:2blockprepot}), (\ref{Habblockdecomp}). For illustration, 
consider the case
\begin{equation}
\label{Hsplit}
H(q_0,q_{\alpha'}) = -\frac{1}{4} \left[
(-q_0) f_1(q_{(n+1)+1}, \ldots, q_{(n+1)+k}) f_2(q_{(n+1)+(k+1)}, \ldots
q_{2n+1}) \right]^{-\frac{1}{2}} \;,
\end{equation}
where the  bottom-right entries 
$\tilde{H}^{\alpha' \beta'}$
split into two sub-blocks of size $k \times k$ and $l\times l$ where 
$k \geq 1$ and $l = n - k$. In this case there are three independent
scalar fields which we can take to be $q_0, q_{(1)} = q_{n+2}, 
q_{(2)} = q_{(n+1)+(k+1)}$. Using the parameters $\propconst_a$ we can
express all charges in terms of three
`independent charges', namely ${\cal Q}_0$ and
\begin{align}
	\mathcal{P}^{(1)} &:= \mathcal{P}^1 = \frac{1}{\propconst_{n+3}} 
\mathcal{P}^2 = \ldots = \frac{1}{\propconst_{(n+1)+k}} \mathcal{P}^k \;, \\
	\mathcal{P}^{(2)} &:= \mathcal{P}^{k +1} = \frac{1}{\propconst_{n+2+k+1}} 
\mathcal{P}^{k+2} = \ldots = \frac{1}{\propconst_{2n+1}} \mathcal{P}^n \;,
\end{align}
where we used that we have chosen $\propconst_{n+2} = \propconst_{(n+1)+(k+1)}=1$.
Note that the solution still depends on all $n+1$ charges
${\cal Q}_0, {\cal P}^A$, which can be chosen freely, but then determine
the ratios between scalar fields belonging to the same block. 
It is however 
convenient to express block-diagonal solutions in terms of charges
${\cal Q}_0, {\cal P}^{(1)}, {\cal P}^{(2)}$ which are in one-to-one
correspondence with the independent scalar fields $q_0, q_{(1)}, q_{(2)}$. 
As we will check below, this system of independent fields and
corresponding charges can be interpreted as a consistent truncation of
the full system.

After eliminating the fields $\dot{\hat{q}}_a$ by their equations
of motion the field equations for the independent scalar fields
are
\begin{align}
		\ddot{q}_{0} - \frac{\left[ \dot{q}_{0}^2 - \mathcal{Q}_{0}^2\right] }{{q}_{0}}  &= 0 \;, \label{BlockNonExtEom1} \\
		\ddot{q}_{(1)} - \frac{\left[ \dot{q}_{(1)}^2 - {\mathcal{P}^{(1)}}{}^2\right] }{{q}_{(1)}}  &= 0 \;, \label{BlockNonExtEom2} \\
		\ddot{q}_{(2)} - \frac{\left[ \dot{q}_{(2)}^2 - {\mathcal{P}^{(2)}}{}^2\right] }{{q}_{(2)}}  &= 0 \;, \label{BlockNonExtEom3} 
	\end{align}
which are solved by
\begin{align}
		q_0 &= \pm \frac{-\mathcal{Q}_0}{B_0} \sinh\left(B_0\tau + B_0 \frac{h_0}{\mathcal{Q}_0} \right) \;, \label{blockklq0soln} \\
		q_{(1)} &= \pm \frac{\mathcal{P}^{(1)}}{B^{(1)}} \sinh\left(B^{(1)}\tau + B^{(1)} \frac{h^{(1)}}{\mathcal{P}^{(1)}} \right) \;, \label{blockklq1soln} \\	
		q_{(2)} &= \pm \frac{\mathcal{P}^{(2)}}{B^{(2)}} \sinh\left(B^{(2)} \tau + B^{(2)}\frac{h^{(2)}}{\mathcal{P}^{(2)}} \right) \label{blockklq2soln} 
\;. 	
\end{align}
The Hamiltonian constraint reduces to 
\begin{equation}
\frac{\left[ \dot{q}_{0}^2 - \mathcal{Q}_{0}^2\right] }{{q}_{0}^2} + \psi_1 \frac{\left[ \dot{q}_{(1)}^2 - \mathcal{P}^{(1)}{}^2\right] }{{q}_{(1)}^2} + \psi_2 \frac{\left[ \dot{q}_{(2)}^2 - \mathcal{P}^{(2)}{}^2\right] }{{q}_{(2)}^2}  = c^2 \;, \label{BlockNonExtEom4}
\end{equation}
in terms of the independent fields, where $\psi_1, \psi_2$ are determined
by the charges through (\ref{psim}). Substituting in the solution, we obtain
\begin{equation}\label{blockklhamconstraint}
	\Big(B_0 \Big)^2 + \psi_1 \left( B^{(1)} \right)^2 + \psi_2 \left(B^{(2)} \right)^2 = c^2 \;.
\end{equation}

\subsubsection{The quantum-deformed $STU$-model}

We conclude this section with a specific example namely the
quantum-deformed $STU$-model with prepotential 
\begin{equation}\label{eq:STU+U^3}
F=-\frac{Y^1Y^2Y^3+a(Y^1)^3}{Y^0} \;.
\end{equation}
This is a particular model where the block $H^{\alpha'\beta'}$ does not
sub-divide, so that we only have the two-block structure of generic
very special prepotentials. While all formulas given in this section
follow straightforwardly from our general results, we give various
formulas explicitly for reference, since this model has many applications.

The Hesse potential for PI configurations is
\begin{equation}
H(u,v) = - 4 \left[ -v_0 \left( u^1u^2u^3 + a(u^1)^3 \right) \right]^{\frac{1}{2}}
\Leftrightarrow 
H(q_a) = - \frac{1}{4}  q_0^{-\frac{1}{2}} \left[ q_5q_6q_7 + a {q_5}^3 \right]^{-\frac{1}{2}} \;.
\end{equation}
This implies
\begin{equation}
\tilde{H}(q_a) \sim \log{q_0} + \log{ \left( q_5q_6q_7+a{q_5}^3 \right)} \;,
\end{equation}
so that from (\ref{tildeHabformula}) we find $\tilde{H}^{ab}$ has the 
following block decomposition 
\begin{equation}
\tilde{H}^{ab} = \frac{\partial^2(-\tilde{H})}{\partial q_a \partial q_b} =\left( \begin{array}{c|cccc|ccc}
\frac{1}{4} q_0^{-2} &  0& 0& 0& 0& 0& 0& 0 \\ \hline \vphantom{\overset{=}{H}} 0  &  *& *& *& *& 0& 0& 0 \\ \vphantom{\overset{=}{H}} 0  & *& *&  *& *& 0& 0& 0 \\ \vphantom{\overset{=}{H}} 0  & *& *& *& *& 0& 0& 0 \\ \vphantom{\overset{=}{H}} 0  & *& *& *& *&  0& 0& 0 \\   \hline \vphantom{\overset{=}{H}} 0 &0 &0&0&0 & \tilde{H}^{55} & \tilde{H}^{56} & \tilde{H}^{57} \\ \vphantom{\overset{=}{H}} 0 &0 &0&0&0 & \tilde{H}^{65} & \tilde{H}^{66} & \tilde{H}^{67}  \\ \vphantom{\overset{=}{H}} 0 &0 &0&0&0 & \tilde{H}^{75} & \tilde{H}^{76} & \tilde{H}^{77}   \end{array} \right) \;.
\end{equation}
We take the independent scalars to be $q_0$ and $q_{(1)}$, where
\begin{equation}
\label{q1}
q_{(1)} := q_5 = \propconst_6^{-1}  q_6 = \propconst_7^{-1} q_7 \;.
\end{equation}
The solution has a full set of $n + 1$ charges ${\cal Q}_0, {\cal P}^A$, and we choose to express the scalar fields in terms of the two charges ${\cal Q}_0$ and
$\mathcal{P}^{(1)} := \mathcal{P}^1$.
The independent fields satisfy the equations of motion
\begin{equation}
\label{eq:stuqeom}
\ddot{q}_0 - \frac{\left[ \dot{q}_0^2 - \mathcal{Q}_0^2 \right]}{q_0} = 0 \;,\;\;\;\ddot{q}_{(1)} - \frac{\left[ \dot{q}_{(1)}^2 - {\mathcal{P}^{(1)}}^2 \right]}{q_{(1)}} = 0 \;,
\end{equation}
with explicit solution 
\begin{align}\label{stu+u3soln}
q_0 &=  \pm \frac{-\mathcal{Q}_0}{B_0} \sinh{ \left( B_0 \tau + B_0 \frac{h_0}{\mathcal{Q}_0} \right)}, \notag
\\ q_{(1)} &= \pm \frac{\mathcal{P}^{(1)}}{B^{(1)}} \sinh{\left( B^{(1)} \tau + B^{(1)} \frac{h^{(1)}}{ \mathcal{P}^{(1)}} \right)} \;.
\end{align}
Substituting this solution into the Hamiltonian constraint
\begin{equation}
\frac{1}{4} \frac{\left[\dot{q}_0^2 - \mathcal{Q}_0^2 \right]}{q_0^2} + \frac{3}{4} \frac{ \left[ \dot{q}_{(1)}^2 - {\mathcal{P}^{(1)}}^2 \right]}{q_{(1)}^2} =c^2 \;.
\end{equation}
gives
\begin{equation}\label{stu+u3hamconstraint}
\frac{1}{4} \Big( B_0 \Big)^2 + \frac{3}{4} \left( B^{(1)} \right)^2 = c^2 \;.
\end{equation}
Observe that the coefficients on the left hand side sum to one. 
As we already remarked below equation $(\ref{SumPsi})$,
we will see in Section \ref{genblockliftsec} that this is a condition 
which is related to the regularity of the lifted four-dimensional solution.

We can use this example to demonstrate that setting scalar fields
belonging to the same block proportional to one another is a 
consistent truncation: if we use
(\ref{q1}) to reduce the Hesse potential to 
\begin{equation}\label{Hstu+u3propscalars}
H(q_0, q_{(1)}) = -\frac{\beta}{2} q_0^{-\frac{1}{2}} q_{(1)}^{-\frac{3}{2}} \;,\;\;\;
\beta =\frac{1}{2} \left( \propconst_6 \propconst_7 + a \right)^{-1/2}\;,
\end{equation}
then using (\ref{tildeHabformula}) we find 
\begin{equation}
\tilde{H}^{00} =  \frac{1}{4}q_0^{-2}, \quad \tilde{H}^{(1)(1)} = \frac{3}{4} q_{(1)}^{-2} \;,
\end{equation}
as well as
\begin{equation}
\tilde{H}_{00} \partial^0 \tilde{H}^{00} = - 2 q_0^{-1} \;,\;\;\;
\tilde{H}_{(1)(1)} \partial^{(1)} \tilde{H}^{(1)(1)} = - 2 q_0^{-1} \;,\;\;\;
\end{equation}
From these relations we obtain the equations of motion  
(\ref{eq:stuqeom}), which thus follow from a one-dimensional
sigma model of the form (\ref{eq:3deffLag}) with Hesse potential 
(\ref{Hstu+u3propscalars}) and the Hamiltonian constraint
(\ref{stu+u3hamconstraint}).

\section{Lifting to four dimensions \label{sec:lifting}}

Having obtained three-dimensional instanton solutions, we now need
to lift them back to four dimensions and identify the subset which 
corresponds to black hole solutions with regular horizons.
Let us therefore explain how one may read off the four-dimensional metric $g^{(4)}_{\hat{\mu} \hat{\nu}}$, gauge fields $F^I_{\hat{\mu} \hat{\nu}}$ and PSK scalar fields $z^A$ from the fields $g^{(3)}_{\mu \nu}, q_a, \hat{q}_a$, which we used to solve the three-dimensional equations of motion in Section \ref{sec:3dsols}. This essentially reverses the dimensional reduction procedure and transformation to dual coordinates given in sections \ref{sec:4dto3d} and \ref{sec:dualcoords}. We will restrict ourselves to spherically symmetric and purely imaginary field configurations of models with prepotentials of the form (\ref{eq:Ff}).

\subsection{General formulas for lifted solutions}

We begin by determining the KK-scalar $e^\phi$ in terms of $q_a$. As seen in (\ref{ephi2H}), this is proportional to the Hesse potential:
\begin{equation}
	e^\phi = -2H\left(q^a(q_b) \right) \;,  \label{eq:Hephi}
\end{equation}
where
\begin{equation}
	H(q^a(q_b)) 
	= -\frac{1}{(2\power+2)} \left[\left(\frac{-q_0}{\power}\right)^\power f \left( q_{n + 2},\ldots, q_{2n + 1} \right) \right]^{-\frac{1}{\power+1}} \;. \label{eq:Hqdual}
\end{equation}
One may then read off the four-dimensional metric $g^{(4)}$ from (\ref{eq:KKdecomp}), using that for static solutions we can set $V_\mu = 0$. Note that the 
three-dimensional part of the metric is fixed to be (\ref{eq:3dmetric}) by the Einstein equations after imposing spherical symmetry. 

We now turn to the gauge fields. First we will need the four-dimensional
complex gauge coupling matrix
\[
{\cal N}_{IJ} = \bar{F}_{IJ} + i \frac{ N_{IK} Y^K N_{JL} Y^L}{Y^M N_{MN}
Y^N} \;,\;\;\;N_{IJ} = 2 \mbox{Im} F_{IJ} \;.
\]
From (\ref{F1st}) we obtain
\[
F_{00}\big|_{PI} = i (-1)^\power \power(\power+1) \frac{f(u)}{(x^0)^{\power+2}} \;,\;\;\;
F_{0A}\big|_{PI} = (-1)^{\power+1} \power \frac{f_A(u)}{(x^0)^{\power+1}} \;,\;\;\;
\]
\[
F_{AB} \big|_{PI} = i (-1)^{\power+1} \frac{f_{AB}(u)}{(x^0)^\power} \;,
\]
which shows in particular that $F_{00}, F_{AB}$ are imaginary while
$F_{0A}$ are real on PI configurations. Next we obtain
\[
N_{00}\big|_{PI} = 2 (-1)^\power \power (\power+1) \frac{f(u)}{(x^0)^{\power+2}} \;,\;\;\;
N_{0A}\big|_{PI} = 0 \;,\;\;\;
N_{AB}\big|_{PI} = 2 (-1)^{\power+1} \frac{f_{AB}(u)}{(x^0)^\power} \;.
\]
Further useful formulae are
\[
(N_{0I} Y^I) \big|_{PI}  = 2 (-1)^\power \power (\power+1) \frac{f(u)}{(x^0)^{\power+1}} \;,\;\;\;
(N_{AI} Y^I) \big|_{PI} = 2 i (-1)^{\power+1} (\power+1) \frac{f_A(u)}{(x^0)^\power} \;,\;\;\;
\]
\[
(Y^I N_{IJ} Y^J)\big|_{PI} = 4 (-1)^\power (\power+1)^2 \frac{f(u)}{(x^0)^\power} \;.
\]
Using these it is straightforward to verify
\[
{\cal N}_{00}\big|_{PI} = i \power (-1)^{\power+1} \frac{f(u)}{(x^0)^{\power+2}} \;,\;\;\;
{\cal N}_{0A} \big|_{PI} = 0 \;,\;\;\;
\]
\[
{\cal N}_{AB} \big|_{PI} = i (-1)^\power \frac{1}{(x^0)^\power}
\left( f_{AB}(u) - \frac{f_A(u) f_B(u)}{f(u)} \right) \;,
\]
which shows in particular that ${\cal N}_{IJ}$ is purely imaginary 
on PI configurations. Note that this does not follow automatically 
from the reality properties that we have imposed. The conditions
by themselves allow real elements 
${\cal N}_{0A}\big|_{PI}$, and it requires an explicit calculation
to see that these matrix elements are in fact zero. 
The actual computation of the four-dimensional
gauge fields is more easily performed using the real version 
\begin{equation}
\label{HhatMatrix}
(\hat{H}_{ab}) = \left( \begin{array}{cc}
{\cal I} + {\cal R}{\cal I}^{-1} {\cal R} & - {\cal R} {\cal I}^{-1} \\
- {\cal I}^{-1} {\cal R} & {\cal I}^{-1} \\
\end{array} \right) \;,\;\;\;{\cal N}_{IJ} = {\cal R}_{IJ}  + i {\cal I}_{IJ}
\end{equation}
of the gauge coupling matrix. As shown above we have ${\cal R}_{IJ}=0$ 
on PI configurations.
The electric components of the four-dimensional gauge fields are
determined by $\dot{\hat{q}}_a = K_a$ to be
\[
\left( \begin{array}{c}
F^I_{t\tau} \\ G_{I|t \tau} \\
\end{array} \right) = 
\left( \begin{array}{c}
- \dot{\zeta}^I \\ - \dot{\tilde{\zeta}}_I \\
\end{array} \right) = - 2 \left( \dot{\hat{q}}^a \right) 
= - 2 \left( \tilde{H}^{ab} K_b  \right) \;.
\]
Using the block structure of $\tilde{H}^{ab}$ as well as
that
\[
K_a = \left( K_0, 0, \ldots, 0, K_{n+2}, \ldots , K_{2n+1} \right) 
= \left(- \mathcal{Q}_0, 0, \ldots, 0,  \mathcal{P}^A \right) 
\]
we obtain:
\[
F^0_{t\tau} = - 2 \tilde{H}^{00} K_0 \;,\;\;\;
F^A_{t\tau} = 0 \;,\;\;\;
G_{0|t \tau} = 0 \;,\;\;\; 
\]
\[
G_{A|t \tau} = -2 \tilde{H}^{A + (n+1), B + (n+1)} K_{B + (n+1)} \;,\;\;\;
A,B=1,\ldots, n \;. 
\]
For PI field configurations, where ${\cal N}_{IJ}$ is 
purely imaginary, the field strength and dual field strength are related by
\[
F^I_{\hat{\mu}\hat{\nu}} = - \frac{i}{2} \sqrt{ | \det g^{(4)} |} 
\epsilon_{\hat{\mu} \hat{\nu} \hat{\rho} \hat{\sigma}}
g^{\hat{\rho} \hat{\alpha}} g^{\hat{\sigma}\hat{\beta}}
{\cal N}^{IJ} G_{J|\hat{\alpha}\hat{\beta}} \;.
\]
Using this relation we can relate the electric components
$G_{I|t\tau}$ of the dual gauge fields to the magnetic components
$F^I_{\theta \phi}$ of the gauge fields. This requires computing
the inverse gauge coupling matrix ${\cal N}^{IJ}$. We use
(\ref{HhatMatrix}) with ${\cal R}_{IJ}=0$ together with the
relation \cite{Mohaupt:2011aa}
\[
\hat{H}_{ab} = H \tilde{H}_{ab} + \frac{2}{H} \Omega_{ac} q^c \Omega_{bd}
q^d \;.
\]
Evaluating this for the block where $a,b, = n+1, \ldots, 2n+1$, 
we obtain:
\[
{\cal N}^{00}\big|_{PI} 
= i \left( H \tilde{H}_{n+1 , n+1} + \frac{2}{H} x^0 x^0\right) \;,\;\;\;
{\cal N}^{0A} \big|_{PI} = 0 \;,\;\;\; \]
\[
{\cal N}^{AB}\big|_{PI} = i H \tilde{H}_{A + (n+1), B + (n+1)} \;.
\]
Using this as well as the explicit form of the four-dimensional
space-time metric $g^{(4)}$ given in (\ref{eq:KKdecomp}) with $V_\mu=0$
and $g^{(3)}$ given by (\ref{eq:3dmetric}) we obtain:
\[
F^0_{\theta \phi} = 0 \;,\;\;\;
F^A_{\theta \phi} = -\frac{1}{2} K_{A+(n+1)} \sin \theta = 
- \frac{1}{2} {\cal P}^A \sin \theta =
- \frac{1}{2} \dot{\hat{q}}_{A+(n+1)} \sin \theta \;.
\]
The results for the field strength can thus be summarised
as
\begin{equation}
\label{eq:4dgauge}
F^0 = -2\tilde{H}^{00} \dot{\hat{q}}_0 dt \wedge d\tau \;,\;\;\;
F^A = -\frac{1}{2} \dot{\hat{q}}_{A+(n+1)} \sin \theta d\theta \wedge
d\phi \;,
\end{equation}
with 
\[
\dot{\hat{q}}_0 = K_0 = - {\cal Q}_0  \;,\;\;\;
\dot{\hat{q}}_{A+(n+1)} = K_{A+(n+1)} = {\cal P}_A  \;.
\]

The complex scalar fields $X^I = e^{-\phi/2} Y^I$ are given by 
\[
X^0 = (-2H)^{-1/2} x^0 \;,\;\;\;
X^A = i (-2H)^{-1/2} u^A \;.
\]
Since we know from Section \ref{SphericalPI} that\footnote{Some care is 
required with regard to signs in the following.
Remember that $H<0$, so that $(H^{\power +1})^{1/(\power +1)} 
= (-1)^\power H$.}
\[
x_0 = \phi_x \left( \frac{\power f(u)}{v_0} \right)^{\frac{1}{\power +1}}
= \phi_x (-1)^\power H \left( \frac{\power f(q_{\alpha'})}{-q_0} 
\right)^{\frac{1}{\power +1}} 
\]
we can express $X^0$ in terms of $q_a$ by
\[
X^0 = \tilde{\phi}_x \sqrt{- \frac{H}{2}} \left( \frac{\power
f(q_{\alpha'})}{-q_0} \right)^{\frac{1}{\power +1}}
\]
where
\[
\tilde{\phi}_x := (-1)^{\power +1} \phi_x= 
\left\{ \begin{array}{ll}
\mbox{sgn}(x^0) & \mbox{for} \;\; \power + 1 \mbox{   even} \\
1 & \mbox{for} \;\; \power + 1 \mbox{   odd} \\
\end{array}
\right. \;,\;\;\;\tilde{\phi}_x^{-1} = \tilde{\phi}_x \;.
\]
Since 
\[
X^A = - i \sqrt{-\frac{H}{2}} q_{A+(n+1)}
\]
the four-dimensional scalars are
\begin{equation}
\label{eq:4dz}
z^A = - i \tilde{\phi}_x q_{A+(n+1)} \left( \frac{-q_0}{\power f(q_{\alpha'})}
\right)^{\frac{1}{\power +1}} \;.
\end{equation}
This is purely imaginary, as required by the PI conditions, provided
that the conditions explained between (\ref{root}) and (\ref{eq:Hreal})
are satisfied. To see this
explicitly, remember how these conditions look in terms of $q_a$:
\begin{itemize}
\item
If $\power+1$ is odd, then there is no sign ambiguity in $x^0$ but
reality of the Kaluza-Klein scalar implies
$f(u) >0 \Leftrightarrow
f(q_{\alpha'}) >0$. In this case $\phi_x = -1$, $\tilde{\phi}_x=1$. 
\item
If $\power +1$ is even, then we need to impose
$v_0 f(u)  > 0 \Leftrightarrow -q_0 f(q_{\alpha'}) >0$.
We can have two different signs: If $v_0>0 \Leftrightarrow q_0>0$,
then $x^0<0$ so that $\tilde{\phi}_x = -1$, whereas if
$v_0<0 \Leftrightarrow q_0 <0$, then $x^0>0$ so that
$\tilde{\phi}_x =1$. 
\end{itemize}
In either case the root is manifestly real, and thus $z^A$ is
manifestly purely imaginary. 
Also note that $z^A$ are homogeneous of degree zero in $q_a$. 

\subsection{Black hole regularity conditions}

Not all four-dimensional solutions 
(\ref{eq:Hephi}) - (\ref{eq:4dz}) obtained by lifting three-dimensional
instanton solutions describe black hole spacetimes. We regard 
four-dimensional solutions to be genuine black holes 
if the following three regularity conditions are met:
	\begin{enumerate}[(i)]
		\item There exists an outer horizon, $E$, of finite area,
		\item The physical (PSK) scalar fields, $z^A$, take finite values on $E$,
		\item The metric is asymptotically Minkowski.

	\end{enumerate}
\noindent The third condition is checked by evaluating $e^\phi$ at radial 
infinity ($\tau \rightarrow 0^+$), and the second by evaluating $z^A$ at the 
horizon $(\tau \rightarrow +\infty)$. For the first condition one must use the formula for the area
\begin{equation}
A= \int_E \mbox{vol}_E  = \lim_{\tau \rightarrow + \infty} \int_{E_\tau} 
\sqrt{ \text{det} g^{(2)}} d \Omega_{(2)} \;,
\end{equation}
where $g^{(2)}$ is the pullback of the four-dimensional metric $g^{(4)}$
to the two-dimensional surface $E_\tau$ given by 
$t, \tau = \text{const}$. $E_\tau$ is independent of $t$, and the
event horizon $E$ is obtained by $\tau \rightarrow +\infty$.

\subsection{Diagonal models}
\label{genliftsec}

We will now turn to explicit examples, starting with 
diagonal models, i.e.\ models with prepotential of the form (\ref{eq:Fsclass}). Three-dimensional instanton solutions were found in the previous section, described by (\ref{eq:qdowninstsoln}) and (\ref{eq:hamconstsub}), which we shall now lift to four dimensions. We may use (\ref{eq:Hephi}) and (\ref{eq:Hqdual}) to write the KK-scalar for this solution as
\begin{equation}\label{eq:warpfactor}
e^\phi = \frac{1}{(\power+1)} \left[\left(\frac{-q_0}{\power}\right)^\power \left( q_{n+2} \dots q_{2n+1} \right)^{\frac{\power+2}{n}} \right]^{-\frac{1}{\power+1}} \;. 
\end{equation}
Using (\ref{eq:KKdecomp}) and (\ref{eq:3dmetric}), we can insert the above warp factor to ascertain the following four-dimensional metric
\begin{align}\label{eq:genclass4dmetric}
	&ds_4^2 = - \frac{1}{(\power+1)} \left[ \left(\frac{-q_0}{\power}\right)^\power  \left( q_{n+2} \dots q_{2n+1} \right)^{\frac{\power+2}{n}}  \right]^{-\frac{1}{\power+1}} dt^2 \; \; +\notag \\
	&\left( \power+1 \right) \left[ \left(\frac{-q_0}{\power}\right)^\power  \left( q_{n+2} \dots q_{2n+1} \right)^{\frac{\power+2}{n}}  \right]^{\frac{1}{\power+1}} \left( \frac{c^4}{\sinh^4{(c \tau)}} d \tau^2 + \frac{c^2}{\sinh^2{(c \tau)}} d \Omega_{(2)}^2 \right) \;. 
\end{align}
The gauge fields are given by 
\begin{align}
	F^0 &= 
	 \frac{\power}{(\power + 1)} \frac{{\cal Q}_0}{q_0^2} dt \wedge d \tau   \;, \qquad
	F^A =  -\frac{1}{2} {\cal P}^A \sin \theta \, d\theta \wedge d\phi \;, \label{eq:diag4dgauge}
\end{align}
and scalar fields by 
\begin{equation}
	z^A = - i \tilde{\phi}_x q_{n+1+A} \left[\left(\frac{-q_0}{\power}\right) \frac{1}{\left( q_{n+2} \dots q_{2n+1} \right)^{\frac{\power+2}{n}}} \right]^{\frac{1}{\power+1}} \;.
\end{equation}
This is the most general stationary field configuration that is spherically symmetric and purely imaginary. Along with the charges ${\cal Q}_0,{\cal P}^1,\ldots, {\cal P}^n$ there are $2n + 2$ further free parameters in this solution: $B_0,B^1,\ldots,B^n$ and $h_0,h^1,\ldots, h^n$, where we interpret $c$ as a dependent parameter, which is determined by (\ref{eq:hamconstsub}). 

We would like to determine for which choices of parameters this solution corresponds to a genuine black hole.
The area of the horizon is given by 
\begin{equation}
A= 4 \pi \lim_{\tau \rightarrow + \infty} \left( \power+1 \right) \left[\left(\frac{-q_0}{\power}\right)^\power   \left( q_{n+2} \dots q_{2n+1} \right)^{\frac{\power+2}{n}} \right]^{\frac{1}{\power+1}} \frac{c^2}{\sinh^2{ \left( c \tau \right)}} \;.
\end{equation}
Expanding the solution of the scalar fields (\ref{eq:qdowninstsoln}) in terms of exponentials tells us that the highest order term in the numerator is
\begin{equation}
\exp \left[ \left( \frac{\power}{\power+1} B_0 + \frac{\power+2}{(\power+1)n} B^1 + \dots + \frac{\power+2}{(\power+1)n} B^n \right) \tau \right] \;. \notag
\end{equation}
Meanwhile, the highest order term in the denominator is given by $e^{2c \tau}$. In order to obtain a finite area these terms must exactly match one-another, which places the following constraint on the integration constants
\begin{equation}\label{eq:finiteA}
\frac{\power}{\power+1} B_0 + \frac{\power+2}{(\power+1)n} B^1 + \dots + \frac{\power+2}{(\power+1)n} B^n = 2c \;.
\end{equation}

We also impose that the physical scalar fields $z^A$ take finite values on the horizon.
In the limit $\tau \rightarrow + \infty$ the $q_a$ scalars behave as $q_0 \sim e^{B_0 \tau}, q_{n+2} \sim e^{B^1 \tau}, \dots, q_{2n+1} \sim e^{B^n \tau}$, and so the only way to guarantee that the $z^A$ remain finite on the horizon is to set
\begin{equation}\label{eq:Bequal}
B_0=B^1= \dots = B^n \;.
\end{equation}
If we combine this with the finite horizon constraint (\ref{eq:finiteA}), we see that $B=c$ i.e. the integration constants satisfy
\begin{equation}
B_0=B^1= \dots = B^n =c \;.
\end{equation}
At this point, we can rewrite the solution in (\ref{eq:qdowninstsoln}) for the scalar fields as

\begin{equation}
\label{eq:constrainqdown}
q_0 = - \frac{\mathcal{Q}_0}{c} \sinh{ \left( c \tau + c \frac{h_0}{\mathcal{Q}_0} \right)} \;,\;\;\;
q_{\alpha'} = \frac{\mathcal{P}^{\alpha'}}{c} \sinh{ \left( c \tau + c 
\frac{h^{\alpha'}}{\mathcal{P}^{\alpha'}} \right)} \;,\;\;
\end{equation}
$\alpha'=n+2, \ldots, 2n+1$.
We also impose that the solution is asymptotically flat. From (\ref{eq:KKdecomp}) we see that in order to obtain Minkowski space at radial infinity we need to ensure that $e^\phi \rightarrow 1$. By (\ref{eq:warpfactor})
this places one more constraint on the integration constants 
\begin{align}
 &\left[ \frac{\mathcal{Q}_0}{\power c} \sinh{ \left(\frac{ch_0}{\mathcal{Q}_0} \right)}\right]^\power \left[ \frac{\mathcal{P}^1}{c} \sinh{ \left( \frac{ch^1}{\mathcal{P}^1} \right)} \dots \frac{\mathcal{P}^n}{c} \sinh{ \left( \frac{ch^n}{\mathcal{P}^n} \right)} \right]^{\frac{\power+2}{n}}  = \left( \power+1 \right)^{-(\power + 1)} \;. \label{eq:AsymMink}
\end{align}

It's worth noting that the constrained scalars above automatically satisfy the Hamiltonian constraint (\ref{eq:hamconstnosub}). The solution described by (\ref{eq:constrainqdown}) satisfies conditions (i),(ii) and (iii) and therefore describes a black hole. Other than the charges, the solution is described by the $n+ 2$ parameters $h_0,h^1,\ldots,h^n$ and $c$. These are subject to one algebraic constraint (\ref{eq:AsymMink}). 
This leaves a total of $n + 1$ independent parameters in the solution 
for $q_a$, $a=0, n+2, \ldots, 2n+1$, which is consistent with and
suggestive of the existence of a first order rewriting of the equations 
of motion. We will come back to this in section \ref{sec:BHIntConsts}.

For the interpretation as a black hole, it is convenient to 
replace the `affine' radial coordinate $\tau$, 
by the radial coordinate $\rho$ defined in (\ref{W}). This rewriting
will make explicit that the four-dimensional metric (\ref{eq:genclass4dmetric}) 
is a deformation of the Reissner-Nordstr\"om metric, and will allow us
to express the solution in terms of harmonic functions.

To demonstrate the rewriting
of the scalar fields, consider $q_0$:
\begin{align}
q_0 &= - \frac{\mathcal{Q}_0}{c} \sinh{ \left( c \tau + c \frac{h_0}{\mathcal{Q}_0} \right)} \notag
\\ &= -  W^{-\frac{1}{2}} \left[ \frac{\mathcal{Q}_0}{c} \sinh{\left( c \frac{h_0}{\mathcal{Q}_0} \right)} + \mathcal{Q}_0 e^{-c \frac{h_0}{\mathcal{Q}_0}} \frac{1}{\rho} \right] \notag
\\ &= \left( \frac{\power^\power}{\left( \power+1 \right)^{\power+1}} \right)^{\frac{1}{2\power+2}} \frac{\mathcal{H}_0}{W^{\frac{1}{2}}} \;,
\end{align}
where
\[
\mathcal{H}_0 = -  \left( \frac{\left( \power+1 \right)^{\power+1}}{\power^\power} \right)^{\frac{1}{2\power+2}} \left[ \frac{\mathcal{Q}_0}{c} \sinh{\left( c \frac{h_0}{\mathcal{Q}_0} \right)} + \mathcal{Q}_0 e^{-c \frac{h_0}{\mathcal{Q}_0}} \frac{1}{\rho} \right]
\] 
is a harmonic function. The prefactors have been chosen such that this and
the following expressions are as simple as possible, while allowing 
$\power$ to be general. Similarly we can express the other scalars
as ratios of harmonic functions:
\begin{equation}
q_{n+1+A} = \left( \frac{\power^\power}{\left(\power+1 \right)^{\power+1}} \right)^{\frac{1}{2\power+2}} \frac{\mathcal{H}^{A}}{W^{\frac{1}{2}}} \;,
\end{equation}
where 
\[
\mathcal{H}^{A} =  \left( \frac{ \left( \power+1 \right)^{\power+1}}{\power^\power} \right)^{\frac{1}{2\power+2}} \left[ \frac{\mathcal{P}^A}{c} \sinh{ \left( c \frac{h^A}{\mathcal{P}^A} \right)} + \mathcal{P}^A e^{-c \frac{h^A}{\mathcal{P}^A}} \frac{1}{\rho} \right] \;.
\] 
The four-dimensional scalar fields can now be expressed in terms of 
the harmonic functions as
\begin{equation}
\label{zHarmonic}
z^A = - i \tilde{\phi}_x \power^{- \frac{1}{\power +1}} \mathcal{H}_A 
\left( \frac{- \mathcal{H}_0}{ (\mathcal{H}^1 \cdots \mathcal{H}^n)^{\frac{
\power +2}{n}} } \right)^{\frac{1}{\power +1}} \;.
\end{equation}

Substituting our results into (\ref{eq:warpfactor}), we find the Kaluza-Klein scalar can be expressed as 
\begin{equation}
e^\phi = \frac{W}{\left[ \left(-\mathcal{H}_0 \right)^\power \left( \mathcal{H}^{1} \dots \mathcal{H}^{n} \right)^{\frac{\power+2}{n}} \right]^{\frac{1}{\power+1}}} 
\;,
\end{equation}
and the four-dimensional metric becomes 
\begin{align}
	ds_4^2 = &- \frac{W}{\left[ \left(-\mathcal{H}_0 \right)^\power \left( \mathcal{H}^1 \dots \mathcal{H}^{n} \right)^{\frac{\power+2}{n}} \right]^{\frac{1}{\power+1}}} dt^2 \notag \\
	&\hspace{4em} + \left[ \left( -\mathcal{H}_0 \right)^\power \left( \mathcal{H}^1 \dots \mathcal{H}^{n} \right)^{\frac{\power+2}{n}} \right]^{\frac{1}{\power+1}} \left( \frac{d \rho^2}{W} + \rho^2 d \Omega_{(2)}^2 \right) \;.
\label{MetricHarmonic}
\end{align}
Writing the metric this way draws parallels with the metric for extremal black 
hole solutions 
which can always be written in terms of harmonic functions,
and which is recovered for $c\rightarrow 0$.
Moreover, the solution for the scalar fields takes, when expressed
in terms of harmonic functions, exactly the same form as for 
extreme solutions. However the coefficients of the harmonic functions
change, and depend on the non-extremality parameter $c$, as made
explicit in (\ref{zHarmonic})

The non-extremal Reissner--Nordstr\"om metric is recovered by 
setting the harmonic functions ${\cal H}_0, {\cal H}^A$ proportional to one-another, in which case
the solution carries $n+1$ independent charges ${\cal Q}_0, {\cal P}^A$, while the scalars $z^A$ are constant. 
This solution corresponds to the universal solution described in section
\ref{sec:universal}
which exists irrespective of a block decomposition.

\subsubsection{$STU$-like models}
\label{STUlikesec}

A one-parameter subclass of diagonal models is given by setting $\power=1$ in (\ref{eq:Fsclass}), resulting in prepotentials of the form 
\begin{equation}
F= \frac{\left(Y^1 \dots Y^n \right)^{\frac{3}{n}}}{Y^0}.
\end{equation}
Since the well-known $STU$ model corresponds to the particular choice 
$n = 3$, we will refer to this class of models as $STU$-like.
For PI field configurations it follows from (\ref{eq:Hqdual}) that 
the Hesse potential takes the form
\[
	H(q_a) = -\frac{1}{4} \left[ -q_0 \left( q_1 \dots q_n \right)^{\frac{3}{n}} \right]^{-\frac{1}{2}} \;.
\]
Explicit expressions for the solution can be obtained 
by substituting $\power=1$ into (\ref{eq:qdowninstsoln}) and (\ref{eq:hatqdot}), with the integration constants subject to the constraint (\ref{eq:hamconstsub}).
The four-dimensional metric is given by
\begin{align}
		ds^2_4 &= - \frac{1}{2} \frac{1}{\sqrt{-q_0 \left(q_{n+1} \ldots q_{2n + 1}\right)^{\frac{3}{n}}}} dt^2 \notag
		\\ &\hspace{4em} + 2\sqrt{-q_0 \left(q_{n+1} \ldots q_{2n + 1}\right)^{\frac{3}{n}}} 
			\left(\frac{c^4}{\sinh^4 c\tau} d\tau^2 + \frac{c^2}{\sinh^2 c\tau} d\Omega^2_{(2)} \right) \;.\notag 
\end{align}
The gauge fields are given by 
\begin{align}
	F^0 &= 
	\frac{1}{2} \frac{\mathcal{Q}_0}{q_0^2} d t \wedge d\tau   \;, \qquad
	F^A = -\frac{1}{2} {\cal P}^A \sin \theta \, d\theta \wedge d\phi \;,\notag
\end{align}
and the scalar fields by 
\begin{equation}
\label{zAdiag}
	z^A = -i\tilde{\phi}_x q_{n+1+A} \sqrt{ \frac{-q_0}{\left( q_{n+2} \dots q_{2n+1} \right)^{\frac{3}{n}}}} \;. 
\end{equation}

For this solution to describe a black hole the scalar fields take the restricted form (\ref{eq:constrainqdown}), and so the number of integration constants  reduces from $2n + 2$ down to $n + 1$. In this case one may re-express the four-dimensional metric in terms of harmonic functions and the isotropic radial coordinate, $\rho$, as 
\begin{align}
ds_4^2 &=- \frac{W}{\sqrt{-\mathcal{H}_0 \left( \mathcal{H}^1 \dots \mathcal{H}^{n} \right)^{\frac{3}{n}}}} dt^2 + \sqrt{-\mathcal{H}_0 \left( \mathcal{H}^1 \dots \mathcal{H}^{n} \right)^{\frac{3}{n}}} \left( \frac{d \rho^2}{W} + \rho^2 d \Omega_{(2)}^2 \right) \;, \notag
\end{align}
where $W$, ${\cal H}_0$ and ${\cal H}^A$ are harmonic functions with respect to the flat metric on ${\mathbb R}^{3}$. We recall that $W=1-\frac{2c}{\rho}=e^{-2c \tau}$ whilst $\mathcal{H}_0, \mathcal{H}^A$ are obtained by substituting $\power=1$ into the expressions in Section \ref{genliftsec}. The four-dimensional scalar fields are given by
\begin{equation}
z^A = - i \tilde{\phi}_x  \mathcal{H}_A 
\sqrt{ \frac{- \mathcal{H}_0}{ (\mathcal{H}^1 \cdots \mathcal{H}^n)^{\frac{
3}{n}} } } \;.
\end{equation}

For $\power +1$ even there are two possible choices for $\tilde{\phi}_x$
in (\ref{zAdiag}).
Note, however, that this does not necessarily imply that we have two
physically inequivalent solutions. The reason is that the sign of
$\tilde{\phi}_x$ can be correlated with that of $q_{(n+1)+A}$. To decide
the sign (and in fact the allowed range) of $z^A$ is a model dependent
problem. Let us illustrate this with the explicit example of the
$STU$ model, where $n=3$. This solution has previous appeared in \cite{Galli:2011fq}.
In this case we know that the manifold parametrised by the physical scalars $z^A$ is isometric to
three copies of the Poincar\'e half plane. Since $u^A=0$, $A=1,2,3$ corresponds to the boundary, we can take
each $u^A$ to be either positive or negative. It is convenient to choose the
same sign for all $u^A$. It is straightforward to verify that 
if we either take all $u^A$ to be positive, or all $u^A$ to be negative, 
the only solutions consistent with all conditions are such that 
$\mbox{Im}z^A <0$, that is all scalars $z^A$ take
values in lower half plane. The standard supergravity fields with 
positive real part are then $S= i z^1$, $T=iz^2$, $U=iz^3$.  
The more conventional description of the  $STU$ model is obtained by including a minus sign 
in the definition of the prepotential. For $F = - (Y^1Y^2Y^3)/Y^0$ one finds by a similar analysis
that taking $u^A >0$ (or $u^A <0$) for all $A=1,2,3$ leads to $\mbox{Im}z^A >0$, and in this case
the standard supergravity fields are $S= -i z^1$, $T=-iz^2$, $U=-iz^3$.

In general the choice of the prepotential determines a range of the scalar fields
where the scalar metric is positive definite. The parameters of a solution 
should then be restricted such that scalar fields take only values within this range.
This analysis is model dependent, and we will not further investigate it in this paper.
The above example illustrates that it is relevant that for $\power$ odd
one has two possible choices for the solution.

\subsubsection{The $F = i\frac{Y^1 Y^2 Y^3 Y^4}{(Y^0)^2}$ model}

Let us next give one explicit example of a diagonal model with 
$\power >1$. For concreteness we choose 
the case $\power=2,n=4$, which is the minimal deviation from the
$STU$ model that is not very special. This model has prepotential
\begin{equation}
F=i \frac{Y^1Y^2Y^3Y^4}{(Y^0)^2} \;,
\end{equation}
and, using (\ref{eq:HPI}), we can show that for PI 
configurations, the Hesse potential assumes the form
\begin{equation}
H(q_a) = - \frac{1}{6} \left[ \frac{1}{4} q_0^2 q_6q_7q_8q_9 \right]^{-\frac{1}{3}} \;.
\end{equation}
Explicit expressions for the solution 
can be found by substituting $\power=2,n=4$ into (\ref{eq:qdowninstsoln}) and (\ref{eq:hatqdot}), with the integration constants constrained by (\ref{eq:hamconstsub}). Dimensionally lifting this solution produces the following four-dimensional metric
\begin{align}\label{s=2n=4metric}
ds_4^2 &= \frac{1}{3} \frac{1}{\sqrt[3]{ \frac{1}{4} q_0^2q_6q_7q_8q_9}} dt^2 \notag
\\ &+ 3 \sqrt[3]{ \frac{1}{4} q_0^2q_6q_7q_8q_9} \left( \frac{c^4}{\sinh^4{(c \tau)}} d \tau^2 + \frac{c^2}{\sinh^2{(c \tau)}} d \Omega_{(2)}^2 \right) \;.
\end{align}
The gauge fields are given by 
\begin{align}
	F^0 &= 
	\frac{2}{3} \frac{\mathcal{Q}_0}{q_0^2} d t \wedge d\tau   \;, \qquad
	F^A = -\frac{1}{2} {\cal P}^A \sin \theta \, d\theta \wedge d\phi \;,\notag
\end{align}
and the scalar fields by 
\begin{equation}
	z^A =- i \tilde{\phi}_x  q_{5+A} \sqrt[3]{ \frac{-q_0}{2 \left( q_6 q_7 q_8 q_9 \right)}} \;. \notag
\end{equation}
As before, for this solution to describe a black hole, the scalar fields must take the restricted form (\ref{eq:constrainqdown}), which again reduces the number of integration constants from $2n+2$ to $n+1$. We can then rewrite the four-dimensional metric in terms of harmonic functions and the isotropic radial coordinate, $\rho$, as
\begin{align}
ds_4^2 &= - \frac{W}{\sqrt[3]{\mathcal{H}_0^2 \mathcal{H}^1 \mathcal{H}^2 \mathcal{H}^3 \mathcal{H}^4}} dt^2 \notag
\\ &+ \sqrt[3]{\mathcal{H}_0^2 \mathcal{H}^1 \mathcal{H}^2 \mathcal{H}^3 \mathcal{H}^4} \left( \frac{d \rho^2}{W} + \rho^2 d \Omega_{(2)}^2 \right) \;,
\end{align}
where $W$ is a harmonic function given by $W=1-\frac{2c}{\rho}=e^{-2 c \tau}$ and $\mathcal{H}_0, \mathcal{H}^A$ are harmonic functions obtained by substituting $\power=2,n=4$ into the expressions in Section \ref{genliftsec}. The four-dimensional scalar fields are
\begin{equation}
z^A = - i \tilde{\phi}_x  \mathcal{H}_A 
\sqrt[3]{ \frac{- \mathcal{H}_0}{ 2(\mathcal{H}^1 \cdots \mathcal{H}^4)} }  \;.
\end{equation}

\subsection{Block diagonal models \label{genblockliftsec}}

We shall now give a description of how to lift the three-dimensional instanton solutions of block diagonal models to four dimensions. 
For concreteness we consider 
the case where the bottom right block decomposes into two sub-blocks; one of size $k \times k$ and one of size $l \times l$ where $k \geq 1$ and $l=n-k$. Instanton solutions to such models were discussed in Section \ref{blockinstantonsec} and are described by (\ref{blockklq0soln}) - (\ref{blockklhamconstraint}). Again, it is possible to use (\ref{eq:Hephi}) and (\ref{eq:Hqdual}) to write the KK-scalar as
\begin{equation}
e^\phi=\frac{1}{2 \sqrt{-q_0 f_1(q_{(1)}) f_2(q_{(2)})}} \;,
\end{equation}
where we have decomposed the function $f$ appearing in (\ref{eq:Hqdual}) as discussed in (\ref{Hsplit}) and set $\power=1$ as the models considered in Section \ref{blockinstantonsec} are all obtainable from five dimensions. We can then use (\ref{eq:KKdecomp}) and (\ref{eq:3dmetric}) to insert this warp factor into the four-dimensional metric as follows
\begin{align}
&ds_4^2 = -\frac{1}{2 \sqrt{-q_0 f_1(q_{(1)}) f_2(q_{(2)})}} dt^2  \notag
\\ &+2 \sqrt{-q_0 f_1(q_{(1)}) f_2(q_{(2)})} \left( \frac{c^4}{\sinh^4{(c \tau)}} d \tau^2 + \frac{c^2}{\sinh{(c \tau)}} d \Omega_{(2)}^2 \right) \;.
\end{align}

From (\ref{eq:4dgauge}), the gauge fields are given by 
\begin{equation}\label{eq:blockgauge}
F^0= \frac{1}{2} \frac{\mathcal{Q}_0}{q_0^2} dt \wedge d \tau, \qquad F^A = -\frac{1}{2} \mathcal{P}^A \sin{\theta} d \theta \wedge d \phi \;,
\end{equation}
and we make the observation that these are exactly the same as for the $STU$-like models considered in Section \ref{STUlikesec} (or indeed any diagonal model with $\power=1$). The only difference is that now the ratios between
scalar fields belonging to the same block are determined by the ratios
of the corresponding charges.

From (\ref{eq:4dz}), the scalar fields assume the form
\begin{equation}
z^A = - i \tilde{\phi}_x q_{A+(n+1)} \sqrt{-\frac{q_0}{f_1(q_{(1)})f_2(q_{(2)})}} \;,
\end{equation}
where $q_{A+(n+1)}$ is proportional to $q_{(1)}$ for $A=1, \ldots, k$ 
and proportional to $q_{(2)}$ for $A=k+1, \ldots, k+l=n$. When viewing
the non-extremality parameter $c$ as being determined by 
(\ref{blockklhamconstraint}), we have, apart from the charges,
$6$ free parameters in the solution: $B_0,B^{(1)},B^{(2)}$ and 
$h_0,h^{(1)},h^{(2)}$.

Even without specifying the functions $f_1$ and $f_2$ we can see
that these 6 parameters reduce to 3 when imposing the conditions
that guarantee a regular black hole solution.
The area of the horizon is
\[
A=8 \pi \lim_{\tau \rightarrow \infty} \sqrt{-q_0 f_1(q_{(1)}) f_2(q_{(2)})} \frac{c^2}{\sinh^2{(c \tau)}} \;.
\]
From (\ref{eq:2blockprepot}), we know that the product $f_1(q_{(1)}) f_2(q_{(2)})$ is homogeneous degree three. Regardless of the individual degrees of homogeneity of $f_1$ and $f_2$, the requirement that the above area be finite together with the requirement that the $z^A$ take finite values on the horizon imply
\[ B_0=B^{(1)}=B^{(2)}=c \;. \]
Moreover, substituting the solution back into the Hamiltonian constraint 
(\ref{BlockNonExtEom4}), we find that this is satisfied provided that
$1+ \psi_1 + \psi_2 =1 \Rightarrow \psi_1 + \psi_2 =0$, using that one of the three blocks in (\ref{blockdecomp2}) only contains
one scalar field $q_0$. For a general decomposition with $M$ blocks one finds that regularity
requires $B^{(m)}=c$ for
$m=1, \ldots, M$ so that the condition becomes (\ref{SumPsi}).

Additionally, the requirement that $e^\phi \rightarrow 1$ as $\tau \rightarrow 0^+$, places one algebraic constraint on the parameters $h_0,h^{(1)},h^{(2)}$. Altogether, these constraints reduce the $6$ free parameters of the instanton solution to $3$ free parameters.

If we know the functions $f_1$ and $f_2$ explicitly, then it is possible to rewrite the metric using the isotropic radial coordinate $\rho$, and with the warp factors being expressed as ratios of harmonic functions.

\subsubsection{The quantum deformed  $STU$ model}

As an explicit example of a block-diagonal model we consider the quantum deformed $STU$ model
with prepotential (\ref{eq:STU+U^3}). 
In this case $\tilde{H}^{\alpha'\beta'}$ does not decompose into
smaller blocks, so that this represents the generic situation for
models obtainable from five dimensions. But we can adapt the
formulae given above by choosing $f_1$ to have degree three and 
$f_2=1$.

The instanton solution is described by two independent scalars as seen in (\ref{stu+u3soln}) and (\ref{stu+u3hamconstraint}). We saw how to write the Hesse potential for such a solution in (\ref{Hstu+u3propscalars}), from which we can use (\ref{eq:Hephi}) to find the KK-scalar is
\begin{equation}\label{stu+u3ephi}
e^\phi = \beta q_0^{-\frac{1}{2}} q_{(1)}^{-\frac{3}{2}} \;,
\end{equation}
where $\beta$ was computed in (\ref{Hstu+u3propscalars}).
We can then substitute this into (\ref{eq:KKdecomp}) to dimensionally lift the instanton solution to the following four dimensional metric
\begin{equation}\label{stu+u3metric}
ds_4^2 = - \beta q_0^{-\frac{1}{2}} q_{(1)}^{-\frac{3}{2}} dt^2 + \frac{1}{\beta q_0^{-\frac{1}{2}} q_{(1)}^{-\frac{3}{2}}} \left( \frac{c^4}{\sinh^4{(c \tau)}} d \tau^2 + \frac{c^2}{\sinh^2{(c \tau)}} d \Omega_{(2)}^2 \right) \;.
\end{equation}
The gauge fields take the same form as in (\ref{eq:blockgauge}) i.e.
\[ F^0= \frac{1}{2} \frac{\mathcal{Q}_0}{q_0^2} dt \wedge d \tau, \qquad F^A = -\frac{1}{2} \mathcal{P}^A \sin{\theta} d \theta \wedge d \phi \;, \]
whilst the scalar fields are
\[ z^A =-2i\tilde{\phi}_x \beta q_{A+4} \sqrt{\frac{ q_0}{q_{(1)}^3}} \;, \;\;\;A=1,2,3\;,\]
where $q_5 = q_{(1)} \propto q_6 \propto q_7$.
For this model, the instanton solution is described (once the ratios of the scalar fields within the
3 by 3 block have been fixed)
by the independent charges $\mathcal{Q}_0, \mathcal{P}^{(1)}$ and the $4$ free parameters $B_0,B^{(1)},h_0$ and $h^{(1)}$. 
The analysis of the conditions required for a regular black hole solution follow from the previous
discussion and leads to the condition 
\begin{equation}
B_0=B^{(1)}=c \;.
\end{equation}
The Hamiltonian constraint (\ref{stu+u3hamconstraint}) is then automatically satisfied.
Asymptotic flatness leads to the further condition
\begin{equation}
\left( -  \frac{\mathcal{Q}_0}{c} \sinh{ \left( c \frac{h_0}{\mathcal{Q}_0} \right)} \right)^{\frac{1}{2}} \left(  \frac{\mathcal{P}^{(1)}}{c} \sinh{\left( c \frac{h^{(1)}}{\mathcal{P}^{(1)}} \right)} \right)^{\frac{3}{2}} = \beta \;,
\end{equation}
on the integration constants $h_0, h^{(1)}$. 

Finally, we can rewrite the four-dimensional metric (\ref{stu+u3metric}) in terms of harmonic functions and the isotropic radial coordinate, $\rho$, as
\begin{equation}
ds_4^2 = - \frac{W}{\mathcal{H}_0^{\frac{1}{2}} {\mathcal{H}^{(1)}}^{\frac{3}{2}}} dt^2 + \mathcal{H}_0^{\frac{1}{2}} {\mathcal{H}^{(1)}}^{\frac{3}{2}} \left( \frac{d \rho^2}{W} + \rho^2 d \Omega_{(2)}^2 \right) \;,
\end{equation}
where the harmonic functions $W, \mathcal{H}_0$ and $\mathcal{H}^{(1)}$ are given by
\begin{align}
W &= 1- \frac{2c}{\rho} = e^{-2c \tau}\;,  \notag
\\ \mathcal{H}_0 &= - \beta^{\frac{1}{2}} \left[ \frac{\mathcal{Q}_0}{c} \sinh{ \left( c \frac{h_0}{\mathcal{Q}_0} \right)} + \mathcal{Q}_0 e^{-c \frac{h_0}{\mathcal{Q}_0}}\frac{1}{\rho} \right] \notag
\\ &= - \beta^{\frac{1}{2}} \left[ \frac{1}{2c} \mathcal{Q}_0 e^{c \frac{h_0}{\mathcal{Q}_0}} - \frac{1}{2c} \mathcal{Q}_0 e^{-c \frac{h_0}{\mathcal{Q}_0}} e^{-2 c \tau} \right] \;, \notag
\\ \mathcal{H}^{(1)} &=  \beta^{\frac{1}{2}} \left[ \frac{\mathcal{P}^{(1)}}{c} \sinh{ \left( c \frac{h^{(1)}}{\mathcal{P}^{(1)}} \right)} + \mathcal{P}^{(1)} e^{-c \frac{h^{(1)}}{\mathcal{P}^{(1)}}} \frac{1}{\rho} \right] \notag
\\ &=  \beta^{\frac{1}{2}} \left[ \frac{1}{2c} \mathcal{P}^{(1)} e^{c \frac{h^{(1)}}{\mathcal{P}^{(1)}}} - \frac{1}{2c} \mathcal{P}^{(1)} e^{-c \frac{h^{(1)}}{\mathcal{P}^{(1)}}} e^{-2 c \tau} \right] \;.
\end{align}
The four-dimensional scalar fields are given by
\begin{equation}
 z^A =-2i\tilde{\phi}_x \beta \propconst_{4+A} {\cal H}^{(1)} \sqrt{\frac{ {\cal H}_0}{{{\cal H}^{(1)}}^3}} \;, \;\;\;A=1,2,3\;,
\end{equation}
where $\propconst_5 = 1,\, \propconst_6 = \dfrac{{\cal P}^2}{{\cal P}^1},\, \propconst_7 = \dfrac{{\cal P}^3}{{\cal P}_1}$, and we therefore have $z^1 \propto z^2 \propto z^3$.

\section{Black holes and first order equations}
\label{sec:BHIntConsts}

For diagonal models, the general solution for spherically symmetric and purely imaginary field configurations (\ref{eq:qdowninstsoln}) satisfies the $n + 1$ first order equations
\begin{equation}
	\dot{q}_\alpha = \sqrt{ (B_\alpha q_\alpha)^2 + K_\alpha^2} \;, \qquad \alpha = 0,n+2,\ldots, 2n + 1\;. \label{eq:FirstOrder}
\end{equation}
Aside from the charges $K_\alpha$, these first order equations contain $n+1$ free parameters $B_\alpha$, which indicate that they have been obtained via integration from second order equations (the equations of motion (\ref{eq:2ndOrderEOM})), and are therefore not unique. There are various different ways one may package the equations (\ref{eq:FirstOrder}), for example
one may write the RHS in terms of $q^\alpha$ coordinates as
\[
	\dot{q}_\alpha = \frac{\sqrt{B'_\alpha{}^2 +  (K_\alpha q^\alpha)^2}}{q^\alpha} \;,
\]
where the $B'_\alpha$ are proportional\footnote{
The exact relations are $B_0' = \tfrac{\power}{2(\power + 1)}B_0$ and  $B_{A + (n + 1)}' = \tfrac{\power + 2}{2n(\power + 1)}B_{A + (n + 1)}$.
}
 to the constants $B_\alpha$. 
One may integrate the RHS to obtain gradient-flow equations $\dot{q}_\alpha = \frac{\partial}{\partial q^\alpha} {\cal W}$, where 
\[
	4 {\cal W} = \sum_\alpha \left[ \sqrt{B'_\alpha{}^2 + (K_\alpha q^\alpha)^2} + \frac{B'_\alpha}{2} \log \left( \frac{\sqrt{B'_\alpha{}^2 + (K_\alpha q^\alpha)^2} - B'_\alpha}{\sqrt{B'_\alpha{}^2 + (K_\alpha q^\alpha)^2} + B'_\alpha} \right) \right] \;.
\]
A similar expression has been previously found for black hole solutions the STU model \cite{Galli:2011fq}, which in our case corresponds to the specific choice $\power = 1, n = 3$. It is worth emphasising that the above gradient flow equations are valid for all solutions to the equations of motion, not just black holes, and that they depend on $n + 1$ free parameters ${\cal W} = {\cal W}(B'_\alpha)$. Therefore the existence of gradient flow equations does not mean that the solution satisfies a unique set of first order equations.

The situation is different for black hole solutions. We have shown that non-extremal black hole solutions are characterised by the requirement that $B_\alpha = c$ for all $\alpha$. In this case (\ref{eq:FirstOrder}) reads
\begin{equation}
	\dot{q}_\alpha = \sqrt{ c^2 q_\alpha^2 + K_\alpha^2} \;. \label{eq:FirstOrder2}
\end{equation}
Aside from the charges, these first-order equations contain just one free parameter: the non-extremality parameter $c$. In other words, we find that black hole solutions \emph{do} satisfy a unique set of $n + 1$ first order  equations that depend only on the charges and non-extremality parameter. In this sense, black hole solutions are characterised by a reduction of the second order equations of motion to first order equations \emph{without increasing the number of equations}. A similar conclusion was also found in our previous investigation into five-dimensional supergravity coupled to vector multiplets \cite{Mohaupt:2012tu}. Counting the number of integration constants, we find that black hole solutions of the scalar fields $q_\alpha$ (or equivalently the complex scalar fields $z^A$, which are purely imaginary, plus the KK-scalar $e^\phi$) contain just $n + 1$ integration constants, compared to $2n + 2$ that are present in the general solution of the equations of motion.
For block-diagonal models the situation is entirely analogous, though in this case we only obtain as many first order equations as there are blocks in the metric.

\section{Conclusion and outlook \label{ConOut}}

In this paper we have continued to develop an approach to non-extremal 
solutions in ${\cal N}=2$ supergravity that is based on the real 
formulation of special geometry,
dimensional reduction over time, and directly solving the second order
field equations. Building upon \cite{Mohaupt:2011aa} we have shown
that non-extremal solutions with one or more non-constant
scalar fields 
can be obtained for a large class
of models by imposing conditions which lead to a block decomposition 
of the equations of motion. Given our ability to a find at least one 
explicit non-trivial solution for each block, we can thus obtain 
explicit solutions, which for the specific conditions we imposed
are given in terms of harmonic functions, as in (\ref{zHarmonic}),
(\ref{MetricHarmonic}). Our method does not rely on group theoretical
methods, and thus is not restricted to homogeneous spaces, nor does
it rely on first order flow equations, and thus allows one to obtain
solutions, and for some models the general solution, to the full
second order equations of motion. 

While we worked with ungauged supergravity and 
used the specific assumptions of spherical symmetry and purely 
imaginary scalar field configurations, it is clear that the method
can be adapted
to various other types of solutions, such as rotating black holes and
black branes, in ungauged supergravity, 
gauged supergravity, and, more general Einstein-Vector-Scalar 
theories with suitable conditions imposed on the couplings. 
We remark that various
features which we can derive and understand systematically within
our formalism have been observed and commented on in the literature
for a variety of models and types of solutions. For example, 
the ansatz for non-extremal solutions which was
recently outlined in \cite{Goldstein:2014qha} relies on various
elements that we have seen at work in the present paper.

One observation commonly shared in the literature is that at least 
some non-extremal solutions
preserve features of BPS solutions. In our work this is manifest when
expressing the solutions in terms of harmonic functions, as in 
(\ref{zHarmonic}), (\ref{MetricHarmonic}):
the line element and gauge fields are modified universally by the additional
harmonic function $W$, while the scalar solution has exactly
the same form as in the BPS case. What changes compared to the BPS case
are the expressions for the constants within the harmonic functions,
which now depend on the non-extremality parameter $c$.

Another universal observation
is that (at least some and maybe all) non-extremal black hole solutions 
satisfy unique first order equations. In our approach this is not an ansatz
or a condition that we impose, but follows when we select from the
general solution of the second order equations the subset that 
describes regular black hole solutions. This reduces the number
of integration constants by one half, and as a result we can
demonstrate that the general black hole solution satisfies a unique set
of first order equations. For BPS and more generally 
extremal solutions the same phenomenon is know to result from
the fixed point behaviour implied by the black hole attractor
mechanism. Since there is no fixed point behaviour for non-extremal
solutions, it is at first surprising that some, and possibly all
non-extremal solutions satisfy first order equations. 
But, as already discussed in \cite{Mohaupt:2009iq,Mohaupt:2010fk,Dempster:2014}
some features commonly associated with the attractor mechanism
persist for non-extremal solutions. 

In fact the synonym `stabilisation
equations' for the BPS attractor equations reflects that obtaining
BPS solutions with regular horizons requires to impose conditions
on the scalar field to `stabilise' them on the horizon. For BPS
solutions this is realised by the  asymptotic restoration 
of full supersymmetry which makes the near horizon solution
a supersymmetric ground state \cite{Gibbons:1981ja,Gibbons:1982fy,Ferrara:1995ih}.
The difference between the extremal and non-extremal case
is that the near horizon solution is a ground state, which forces
the scalars to take fixed point values which are exclusively 
determined by the electric and magnetic charges. In the non-extremal
case the scalar flow reaches the horizon before reaching a fixed
point, and the horizon values of the scalar are not determined
by the gauge charges. But they are still not independent 
integration constants, as they would be if we considered the
full second order scalar equations without regularity conditions
at the horizon. Instead they are determined by other integration constants,
namely the gauge charges together with the asymptotic values of
the scalars at infinity. 
It is therefore not unreasonable to 
expect that the scalar flow between infinity and horizon is 
always governed by first order equations which result from 
deforming the first order equations valid for the extremal case.
While we have demonstrated this here for a large class of models,
it remains to investigate whether this is true in general. 

In the case that the target manifold is a Riemannian symmetric space it has previously been 
observed that the coefficient of the leading order term in the $1/\rho$ expansion of the scalar fields, referred to as 
the scalar charge, is not an independent parameter for black holes solutions \cite{Breitenlohner:1987dg}. In this paper we have constructed full analytic solutions to the equations of motion,
and therefore the reduction in the number of free parameters in the solution is a stronger statement, even when the target manifold is a symmetric space. In fact, by considering the $1/\rho$ expansion one automatically finds that the scalar charge is not an independent parameter, regardless of whether or not the target manifold is symmetric.

While we have only obtained the general solution to the second
order equations of motion for diagonal models, these form
a large class of models with, 
up to two exceptions, non-homogeneous target spaces.
Moreover we saw that the observed pattern persisted for 
block-diagonal models, where we could obtain a subset of
solutions to the second order equations and still observe that
regularity at the horizon reduces the number of integration 
constants by one half. We do not see any reason why these
systematic features should only apply to models where we can
solve the equations of motion explicitly, and expect that
they are generic.

Another universal feature, which for example has also been mentioned
recently in \cite{Goldstein:2014qha} is that our ability to find
explicit
non-extremal solutions results from a symmetry of the equations of motion. 
In our case the relevant symmetry only comes into existence after consistently
truncating out half of the scalars by the PI conditions. The resulting
block decomposition of the Hessian metric implies an invariance
of the equations of motion under a field rotation matrix, which
was discussed in detail in \cite{Mohaupt:2011aa} for the
special case of prepotentials with $\power=1$. We note that this
symmetry can always be used for both generating non-BPS extremal
solutions from BPS solutions (as done in \cite{Mohaupt:2011aa}),
and to obtain non-extremal solutions (as done in the present paper).
This is complementary to the
observation that BPS and non-BPS extremal solutions can be `unified'
through obtaining them both as limits of non-extremal solutions
\cite{Galli:2011fq}.

There are various directions to be explored in the future. As already
mentioned the formalism developed here can be extended and adapted to
gauged supergravity and other types of solutions.
It would be interesting to find situations where
a block decomposition is possible but the solutions for individual
blocks are not harmonic functions. For example, some multi-centered
extremal solutions found for symmetric target spaces contain 
non-harmonic functions \cite{Bossard:2013oga}, and so-called
unconventional solutions involving anharmonic terms were 
constructed in \cite{Bueno:2013pja}.

One limitation of the PI 
condition is that eliminates half of the charges and at least 
half of the independent scalar fields. This was necessary in order
to obtain a block decomposition and to get rid of the terms
in the second and third line of (\ref{L3qupstairs}). However for
some models with symmetric target spaces solutions with all
charges turned on are known, and for the $STU$ model the general
charged rotating solution (including NUT charge) was found in 
\cite{Chow:2013tia}.
It would be interesting to
obtain solutions with more charges turned on 
for non-symmetric and in fact non-homogeneous
target spaces.

While in this paper we have focused on obtaining explicit solutions
in closed form, there is a complementary, more geometrical approach 
about which we will report elsewhere \cite{Cortes:2014,Submanifolds}.
The target manifold of the three-dimensional Euclidean theory
is a para-quaternionic K\"ahler manifold (as is proved in generality 
in \cite{Vaughan:2012,Cortes:2014}), and the construction of solutions
is facilitated by constructing harmonic maps onto totally geodesic
submanifolds. The submanifold corresponding to static, purely imaginary
field configurations is in fact a para-K\"ahler submanifold, which
contains the `black string submanifold' already identified in
\cite{Dempster:2013mva}. 
Further para-K\"ahler submanifolds
can be constructed systematically \cite{Dempster:2014,Submanifolds}.
One interesting question for the future is to relate this approach to
the group-theoretical approach which works so well if the target space
is a symmetric space. This will hopefully lead to further insights which
will allow us to obtain a systematic understanding of non-extremal
(and also of extremal) solutions for generic ${\cal N}=2$ string
compactifications.

In this paper we have mostly restricted our attention to those solutions which
are regular four-dimensional black holes. Within the full class of solutions we 
constructed, there should be interesting subclasses corresponding to 
three-dimensional instanton solutions with finite action, and to 
black hole solutions that are only regular when lifting to dimensions
higher than four. We leave this investigation to future work.

\subsubsection*{Acknowledgements} 

The work of T.M. is supported in part by STFC  
grant ST/G00062X/1. The work of D.E. is supported
by STFC studentship ST/K502145/1. The work of O.V. is supported by the German Science Foundation
(DFG) under the Collaborative Research Center (SFB) 676 ``Particles, Strings
and the Early Universe." 
T.M. thanks the 
Department of Mathematics at the University of Hamburg
for hospitality and support during various stages of this work.

\appendix

\section{Hessian geometry \label{App:Hessian}}

\subsection{The Hesse potential $H$}

In this appendix we collect or prove certain identities for Hessian
metrics which we use in the paper or find generally noteworthy.

In terms of affine coordinates $q^a$ a Hessian metric $H_{ab}$ is given by the
second derivatives of a real valued function, the Hesse potential $H$
\[
H_{ab} = \frac{\partial^2 H}{\partial q^a \partial q^b} \;.
\]
The coordinate-independent definition requires the existence of a 
flat, torsion free connection $\nabla$, such that the rank three
tensor $\nabla g$, where $g$ is the metric, is totally symmetric
\cite{2008arXiv0811.1658A}. 
The affine coordinates $q^a$ are then defined by $\nabla dq^a=0$. 

Affine special K\"ahler (ASK) manifolds are simultaneously 
K\"ahler and Hessian.\footnote{We are using the formulation of
special geometry developed in \cite{Alekseevsky:1999ts}. The relevant
facts are reviewed in \cite{Mohaupt:2011aa}.}
One can choose special real coordinates $q^a$ which are affine coordinates
with respect to the Hessian structure and simultaneously
Darboux coordinates, that is the K\"ahler form is constant in these 
coordinates \cite{Freed:1997dp}:
\[
\omega = \Omega_{ab} dq^a \wedge dq^b \;, \;\;\;
(\Omega_{ab})=
\left( \begin{array}{cc}
0 & \mathbbm{1} \\
-\mathbbm{1} & 0 \\
\end{array} \right) \;.
\]
The associated complex structure is
\[
J^a_{\;\;c} = -\frac{1}{2} \Omega^{ab} H_{bc} \;.
\]
It is useful to note the equivalent relation
\[
H_{ab} \Omega^{bc} H_{cd} = - 4 \Omega_{ab} \;.
\]
In ${\cal N}=2$ supergravity the Hesse potential $H$ is homogeneous
of degree 2, which implies the relations
\[
q^a H_a = 2 H \;,\;\;\;
q^a H_{ab} = H_b \;,\;\;\;
q^a H_{abc} = 0 \;,
\]
where $H_a = \frac{\partial H}{\partial q^a}$, etc. This implies
that the affine special K\"ahler manifold is conical, see
\cite{Cortes:2009cs} for a coordinate-free definition. 
While in general the Hesse potential is only unique up to
affine transformations, preserving homogeneity restricts this 
to linear transformations, and chooses affine coordinates which
are adapted to the conical structure. 
Moreover, for special real coordinates
one also imposes that the K\"ahler form is invariant, which further
restricts the linear transformations to be symplectic. 
In the following it is
understood that we use special coordinates which are adapted 
to the conical structure.

Affine special K\"ahler manifolds come in fact equipped with 
a one-param\-eter family of special connections $\nabla$, each with its
own system of special real coordinates \cite{Alekseevsky:1999ts}. 
In particular, dual special real coordinates are defined by
\[
q'_a = H_a = \frac{\partial H}{\partial q^a} \;.
\]
Since $H$ is homogeneous of degree 2, 
the special coordinates and dual special coordinates 
are related by
\[
q'_a = H_{ab} q^b \Leftrightarrow q^q = H^{ab} q'_b\;,
\]
where $H^{ab}$ denotes the inverse of the Hessian metric $H_{ab}$. 
Since $q'_a$ are special real coordinates, and the metric is
\[
g = H_{ab} dq^a dq^b = H^{ab} dq'_a dq'_b \;,
\]
there exists a Hesse potential $H'(q'_a)$ for the inverse
metric $H^{ab}$:
\[
H^{ab} = \frac{\partial^2 H'}{\partial q'_a \partial q'_b} \;.
\]
We now show that corresponding Hesse potential $H'(q')$ 
is given by transforming $H(q)$ with the diffeomorphism
$q^a \mapsto q'_a$, that is $H'(q') = H(q(q'))$. 
Note that the  diffeomorphism $q^a \mapsto q'_a$ is in general 
non-linear (unless $H_{ab}$ is constant), and therefore does not 
preserve the affine structure determined by a given fixed special 
connection $\nabla$. As already mentioned $q^a$ and
$q'_a$ are special real coordinates with respect to two different
affine structures, and in particular need not be related  by a symplectic
transformation.

\textbf{Begin proof}
First note that 
\[
\frac{\partial q'_a}{\partial q^b} = H_{ab} \;,\;\;\;
\frac{\partial q^a}{\partial q'_b} = H^{ab} \;.\;\;\;
\]
Since $\partial q'_a/\partial q^b$ is the Jacobian of
the transformation $q^a \mapsto q'_a$, it is clear that
the metric coefficients with respect to the coordinates $q'_a$ 
are the inverse $H^{ab}$ of the metric coefficients $H_{ab}$
with respect to $q^a$. Since $q'_a$ are special real coordinates,
there exists as a Hesse potential $H'(q')$ 
\[
H^{ab} = \frac{\partial^2 H'}{\partial q'_a \partial q'_b}  \;,
\]
which is homogeneous of degree two.
Our claim is that $H'(q')$ is related to $H(q)$ by
$H'(q') = H(q(q'))$. Since this is equivalent to
$H(q)= H'(q'(q))$, we can prove instead that $H'(q'(q))$ 
is a Hesse potential for $H_{ab}$, i.e. 
\[
H_{ab} = \frac{\partial^2 H'}{\partial q^a \partial q^b} \;.
\]
Using the chain rule, we compute
\[
\frac{\partial^s H'}{\partial q^a \partial q^b}  =
\frac{\partial^2 H'}{\partial q'_c \partial q'_d}
\frac{\partial q'_c}{\partial q^a}
\frac{\partial q'_d}{\partial q^b} 
+
\frac{\partial H'}{\partial q'_c} 
\frac{\partial^2 q'_c}{\partial q^a \partial q^b} 
= H^{cd} H_{ca} H_{db} + \frac{\partial H'}{\partial q'_c} H_{abc}\;.
\]
Then it remains to show that the second term is zero. 
We note that
\[
\frac{\partial^2 H'}{\partial q'_a \partial q'_b} = H^{ab} 
= \frac{\partial q^a}{\partial q'_b} \;,
\]
which can be integrated to
\[
\frac{\partial H'}{\partial q'_a} = q^a \;.
\]
Note that there is no integration constant since $H'$ is
homogeneous of degree two. Using homogeneity we find
\[
\frac{\partial H'}{\partial q'_c} H_{abc} =
q^c H_{abc} = 0 \;,
\]
so that
\[
\frac{\partial^2 H'}{\partial q^a \partial q^b}  =
H^{cd} H_{ca} H_{db} = H_{ab} \;.
\]
\textbf{End proof}

\subsection{The Hesse potential $\tilde{H}$}
\label{dualqsec}
\label{sec:dualcoords}

Given a Hesse potential $H(q)$ which is homogeneous of degree 
two,\footnote{Generalising the following discussion to the case where 
$H(q)$ has an arbitrary degree of homogeneity is straightforward and 
only changes some numerical coefficients in the formulae given in this
section. In five dimensions one can consider non-supersymmetric theories
based on `generalised special real geometry', and it turns out that
black brane solutions can be constructed by the same methods as used
in supergravity \cite{Mohaupt:2009iq,Mohaupt:2012tu,Dempster:2013mva}.
Here we focus on the case of degree two for concreteness, and because
it is the case we consider in this paper.}
we can define a new Hesse potential by 
\begin{equation}\label{CtildeHformula}
\tilde{H} = C \log H \;,
\end{equation}
where $C$ is a constant. In the main part of the paper, we have to choose 
$C=-\frac{1}{2}$ in order for (\ref{tildeHformula}) to hold true. 
Note that if we replace $H$ by $\alpha H$ in (\ref{CtildeHformula}),
where $\alpha \in \mathbbm{R} \backslash \{0\}$, all derivatives remain
unchanged and, since it is derivatives of $\tilde{H}$ that appear in the equations of motion, we are free to make such a change. A constant $\alpha <0$ is 
for example required if $H<0$ in order that the argument of the logarithm is positive. In the main part of the paper we choose $\alpha=-2$, as in 
(\ref{tildeHformula}), because this is convenient when
imposing the D-gauge condition (\ref{ephi2H}).

Returning to our analysis of the general formula (\ref{CtildeHformula}), we see that while $\tilde{H}$ is not a homogeneous function, its $n$-th derivative
is homogeneous of degree $-n$ for $n\geq 1$. 
Therefore the Hessian metric defined by
\[
\tilde{H}_{ab} = \frac{\partial^2 \tilde{H}}{\partial q^a \partial q^b}
\]
has metric coefficients which are homogeneous of degree $-2$, while
the metric tensor $\tilde{g} = \tilde{H}_{ab} dq^a dq^b$ is homogeneous of
degree 0. The metric coefficients can be expressed in terms of $H$ by
\[
\tilde{H}_{ab} = C \frac{H_{ab} H - H_a H_b}{H^2}  \;.
\]
Using homogeneity, it is straightforward to verify that 
the inverse metric has coefficients
\begin{equation}
\label{Htildeinv}
\tilde{H}^{ab} = C^{-1} ( H H^{ab} - q^a q^b) \;.
\end{equation}

We define dual coordinates with respect to 
$\tilde{H}$ by
\begin{equation}
\label{dualtilde}
q_a := \tilde{H}_a := \frac{\partial \tilde{H}}{\partial q^a} =
C \frac{H_a}{H} = C \frac{q'_a}{H}  \;.
\end{equation}
Then
\[
\tilde{H}_{ab} = \frac{\partial q_a }{\partial q^b}
\Rightarrow \tilde{H}^{ab} = \frac{\partial q^a}{\partial q_b} \;.
\]
Since $\tilde{H}_a$ is homogeneous of degree $-1$:
\begin{equation}\label{qupqdown}
\tilde{H}_{ab} q^b = - \tilde{H}_a = - q_a \Rightarrow
q^a = - \tilde{H}^{ab}  q_b \;.
\end{equation}
Due to the additional minus sign, the coordinates
$q^a$ and $q_a$ are not simply related by 
`lowering the index' using the metric $\tilde{H}_{ab}$.
Since coordinates are functions, and not vector fields on the
underlying manifold $M$, there is nothing wrong with this relation.
We do of course
observe the standard tensorial behaviour when considering the action of the metric on tensors, such as tangent vectors to curves
\begin{equation}\label{tildeHderivativereln}
\dot{q}^a = \tilde{H}^{ab} \dot{q}_b \;,
\end{equation} 
partial derivatives\footnote{Since $q^a$ are affine coordinate with
respect to the flat, torsion-free connection $\nabla$ defining the
Hessian structure, partial derivatives
coincide with covariant derivatives in this coordinate system, and 
hence define a covariant object.}
$\frac{\partial}{\partial q^q}$ and
differentials like $dq^a$. 

We can define a dual Hesse potential $\tilde{H}'(q_b):=
\tilde{H}(q^a(q_b))$. 
Then
\[
\frac{\partial \tilde{H}'}{\partial q_a} = 
\frac{\partial  \tilde{H}}{\partial q^b} \frac{\partial q^b}{\partial q_a}
= \frac{\partial  \tilde{H}}{\partial q^b} \tilde{H}^{ba}\;,
\]
which implies that
\begin{equation}
\label{qa}
q^a = - \tilde{H}^{ab} {q}_b =
- \tilde{H}^{ab} \frac{\partial \tilde{H}}{\partial q^b} 
= -\frac{\partial \tilde{H}'}{\partial q_a} \;.
\end{equation}
Therefore $-\tilde{H}'$ is a Hesse potential
for the inverse $\tilde{H}^{ab}$ of $\tilde{H}_{ab}$:
\begin{equation}
\label{Htildeinv2}
\tilde{H}^{ab} = \frac{\partial q^a}{\partial q_b}
= \frac{\partial^2 ( - \tilde{H}')}{\partial q_a \partial q_b}  \;.
\end{equation}

We add some useful relations 
between the two types of dual coordinates, $q'_a = H_a$ and 
$q_a = \tilde{H}_a$. 
From the definition (\ref{dualtilde}) of $q_a$ we derive
\[
\frac{\partial q_a}{\partial q'_b} = C \frac{ \delta_a^b H -
q'_a q^b}{H^2}  \Rightarrow 
\frac{\partial q'_a}{\partial q_b} 
= C^{-1} \left( \delta_a^b H - q'_a q^b \right) \;.
\]
Using this,  one can derive the relations
(\ref{qa}), ({\ref{Htildeinv}}) and (\ref{Htildeinv2})
directly by differentiating the dual Hesse potential 
$\tilde{H}'(q_a) = C \log  H'(q'(q_a)$:
\[
\frac{\partial \tilde{H}'}{\partial q_a} = - q^a \;,\;\;\;
\frac{\partial^2 \tilde{H}'}{\partial q_a \partial q_b}
= - C^{-1} \left( H^{ab} H - q^a q^b \right) = - \tilde{H}^{ab} \;.
\]
In the paper we  compute $\tilde{H}^{ab}$ by
(\ref{Htildeinv2}) with $\tilde{H}' = C \log H''$ where
$H''(q_b) = H(q^a(q_b))$:
\begin{equation}
\tilde{H}^{ab} = C \left( \frac{1}{H''} \frac{\partial^2 H''}{\partial q_a
\partial q_b} - \frac{1}{H^2} \frac{\partial H''}{\partial q_a}
\frac{\partial H''}{\partial q_b} \right) \;.
\label{tildeHabformula1}
\end{equation}
Using the Jacobian $\frac{\partial q_a}{\partial q'_b}$ given above
it is straightforward to check that this is related to 
(\ref{Htildeinv}) by a change of variables.

For notational simplicity, we have usually dropped the primes on 
$\tilde{H}', H', H''$ in the main part of the paper, whenever it is clear 
from context which variables the function depends on.

\section{Spherically symmetric metrics \label{SphericalSym}}

Here we will review material from \cite{Hawking:1973uf,Wald:1984rg}.

A spacetime is said to be spherically symmetric if the isometry group contains a subgroup isomorphic to SO$(3)$, and the orbits of this subgroup are two-spheres. We may therefore interpret SO$(3)$ transformations as rotations. 

The spacetime metric induces a metric on each orbit two-sphere. Since the orbits are two-dimensional submanifolds, and a three-dimensional isometry group is the maximum possible,%
\footnote{The data needed to describe a Killing vector at a point are $\xi^\mu$ and $\nabla_{[\mu}\xi_{\nu]}$. This is because all higher derivatives are determined by the Riemann curvature tensor through
\[
	\nabla_\mu \nabla_\nu \xi_\rho = R_{\mu \nu \rho}^{\phantom{\mu \nu \rho} \sigma} \xi_\sigma \;.
\]
We therefore count $d$ independent degrees of freedom from $\xi^\mu$, and $(d-1)/2$ from $\nabla_{[\mu}\xi_{\nu]}$. The maximum number of Killing vectors is therefore $d(d+1)/2$.}
 $3 = 2(2+1)/2$, the curvature of the two-spheres must be constant. The metric on the orbit two-spheres must therefore be proportional to the metric on the unit two-sphere. By theorem 3 of \cite{Schmidt:1967} at each point the orbit two-spheres are orthogonal to a two-dimensional timelike submanifold, which we parametrise by $(r,t)$. The spacetime metric therefore decomposes into two blocks
\begin{equation}
	ds^2 = \left[-A^2(r,t) dt^2 + B(r,t)dt dr + C^2(r,t) dr^2\right] + D^2(r,t) \left( d\theta^2 + \sin^2 \theta d \varphi^2 \right) \;. \label{eq:spherical}
\end{equation}
In these coordinates a basis of SO$(3)$ rotations is given by
\begin{align*}
	\eta_1 &= -\cos \varphi \, \partial_\theta + \cot \theta \sin \varphi \, \partial_\varphi \;, \\
	\eta_2 &= \sin \varphi \, \partial_\theta + \cot \theta \cos \varphi \, \partial_\varphi  \;, \\
	\eta_3 &= \partial_\varphi\;.
\end{align*}

\subsection{Stationary and spherically symmetric \label{app:A}}

A spacetime is said to be stationary if the isometry group contains a one-parameter subgroup with orbits given by timelike curves, which we paramet\-rise by $t$. This is equivalent to the existence of a timelike Killing vector field $\xi = \partial_t$, which we assume to be unique.

Consider a spacetime that is both stationary and spherically symmetric. Due to the uniqueness of $\xi$ it is orthogonal to the SO$(3)$ orbit two-spheres \cite{Wald:1984rg}. This means that the decomposition of the metric according to (\ref{eq:spherical}) is compatible with the choice of $t$ as a timelike coordinate, and since $t$ parametrises an isometry the components of the metric must be independent of this parameter
\[
	ds^2 = \left[-A(r)^2 dt^2 + B(r)dt dr + C(r)^2 dr^2\right] + D^2(r) \left( d\theta^2 + \sin^2 \theta d \varphi^2 \right) \;.
\]

Let us investigate the function $D$ further. Setting $D$ to be constant, i.e.\ $\nabla_\mu D = 0 \; (= \partial_r D)$, is inconsistent with the equations of motion for either a vacuum solution or a static perfect fluid solution \cite{Hawking:1973uf}, and is therefore not considered physical. We therefore assume $\nabla_\mu D \neq 0$, and we may use the function $D$ as a spacetime coordinate 
\[
	\tilde{r} := D(r) \;, \qquad \frac{d\tilde{r}}{dr} = \partial_r D \neq 0 \;,
\]
in which case the metric takes the form
\[
	ds^2 = \left[-\tilde{A}(\tilde{r})^2 dt^2 + \tilde{B}(\tilde{r})dt d\tilde{r} + \tilde{C}(\tilde{r})^2 d\tilde{r}^2\right] + \tilde{r}^2 \left( d\theta^2 + \sin^2 \theta d \varphi^2 \right) \;.
\]

We shall now review the argument that a stationary and spherically symmetric spacetime is necessarily static. 
First note that 
\begin{equation}
	\nabla_\xi \tilde{r} = \nabla_\xi D(r) = \partial_t D(r) = 0 \;. \notag
\end{equation}
From this expression we can see that the covectors $g(\xi, \cdot)$ and $\nabla \tilde{r} = d\tilde{r}$ are orthogonal, which means that the corresponding vectors
\[
 \xi = \partial_t \qquad\text{and} \qquad  \psi = (\nabla^\mu\tilde{r}) \partial_\mu 
\]
 are orthogonal since they are obtained by raising indices using the metric.
Using the decomposition of the metric we can write $\psi$ as
\begin{align*}
	\psi &=  g^{\tilde{r}t} \frac{\partial}{\partial t} + g^{\tilde{r}\tilde{r}} \frac{\partial}{\partial \tilde{r}} \\
	&=  \frac{-1}{{\tilde{A}}^2 {\tilde{C}}^2 + \tfrac14 {\tilde{B}}^2} \left( -\tfrac12 \tilde{B} \frac{\partial}{\partial t} - \tilde{A}^2 \frac{\partial}{\partial \tilde{r}} \right) \;.
\end{align*}
Since $\xi$ and $\psi$ are orthogonal it follows that $\tilde{B} = 0$. 
It is also clear that $\psi,\,\partial_\theta,\,\partial_\varphi$ commute, and therefore define a three-dimensional integrable distribution. Let us denote the corresponding hypersurface by $\Sigma$, which may be locally parametrised by $(\tilde{r},\theta, \varphi)$. Since $\Sigma$ is orthogonal to the timelike Killing vector $\xi$ we have proved that the spacetime is static. The metric takes the form
\[
	ds^2 = -\tilde{A}(\tilde{r})^2 dt^2 + \tilde{C}(\tilde{r})^2 dr^2 + \tilde{r}^2 \left( d\theta^2 + \sin^2 \theta d \varphi^2 \right) \;.
\]

We end by making the coordinate transformation $\tau = \int \frac{\tilde{C}(\tilde{r})}{\tilde{r}^2 \tilde{A}(\tilde{r})} d\tilde{r}$, in which case the metric may be written as
\begin{equation}
	ds^2 = -\tilde{A}(\tau)^2 dt^2 + \tilde{A}(\tau)^{-2} \left[ e^{4{\cal A}(\tau)} d\tau^2 + e^{2 {\cal A}(\tau)}  \left( d\theta^2 + \sin^2 \theta d \varphi^2 \right) \right] \;,
	\label{eq:MetricTau}
\end{equation}
where $e^{{\cal A}(\tau)} := \tilde{r}\tilde{A}(\tilde{r})$. The advantage of this parametrisation is that 
$\Delta f = \frac{d^2}{ d\tau^2} f$ for all functions $f = f(\tau)$ that only depend on the radial coordinate, which leads to simplifications in the
equations of motion. In particular $\tau$ provides an affine parametrisation 
of the geodesic curve 
$(q^a(\tau),\hat{q}^a(\tau))$ on the scalar manifold 
corresponding to the solution of the
scalar field equations.

\providecommand{\href}[2]{#2}\begingroup\raggedright\endgroup

\end{document}